\documentclass[12pt,a4paper]{article}

\usepackage{epsfig}
\usepackage{amsmath,amsfonts,amssymb}
\usepackage{cite}
\usepackage{colortbl}
\usepackage{pifont}

\newcommand{\be}{\begin{equation}}
\newcommand{\ee}{\end{equation}}

\newcommand{\bea}{\begin{eqnarray}}
\newcommand{\eea}{\end{eqnarray}}
\newcommand{\gsim}{\lower.7ex\hbox{$\;\stackrel{\textstyle>}{\sim}\;$}}
\newcommand{\lsim}{\lower.7ex\hbox{$\;\stackrel{\textstyle<}{\sim}\;$}}
\def\simlt{\stackrel{<}{{}_\sim}}

\providecommand{\openone}{\leavevmode\hbox{\small1\kern-3.8pt\normalsize1}}
\newcommand{\oh}{{\textstyle \frac{1}{2}}}

\newcommand{\Smll}{\Sigma m_{\ell \ell}}

\begin{document}

\vspace*{-2.5cm}
\begin{flushright}
IFT-UAM/CSIC-19-142 \\
\end{flushright}
\vspace{0.cm}

\begin{center}
\begin{Large}
{\bf Multilepton dark matter signals}
\end{Large}

\vspace{0.5cm}
J.~A.~Aguilar--Saavedra$^{a,b}$, J.~A. Casas$^{a,c}$, J. Quilis$^{a,d}$, R. Ruiz de Austri$^e$ \\[1mm]
%\end{center}
\begin{small}
{$^a$ Instituto de F\'isica Te\'orica, IFT- UAM/CSIC, Univ. Aut\'onoma de Madrid, E-28049 Madrid, Spain} \\ 
{$^b$ Universidad de Granada, E-18071 Granada, Spain (on leave)} \\ 
{$^c$ Dept.  of  Physics,  Univ.  of  Notre  Dame, 225  Nieuwland  Hall,  Notre  Dame,  IN  46556,  USA}\\
{$^d$ Fundaci\'on CIEN, Instituto de Salud Carlos III, Fundaci\'on Reina Sof\'\i a, Valderrebollo 5, E-28031 Madrid, Spain} \\
 {$^e$ Instituto de F\'{\i}sica Corpuscular, IFIC-UV/CSIC, Universitat de Val\`encia, E-46890  Paterna, Spain}
\end{small}
\end{center}

\begin{abstract}
The signatures of dark matter at the LHC commonly involve, in simplified scenarios, the production of a single particle plus large missing energy, from the undetected dark matter. However, in $Z'$-portal scenarios anomaly cancellation requires the presence of extra {\em dark leptons} in the dark sector. We investigate the signatures of the minimal scenarios of this kind, which involve cascade decays of the extra $Z'$ boson into the dark leptons, identifying a four-lepton signal as the most promising one. We estimate the sensitivity to this signal at the LHC, the high-luminosity LHC upgrade, a possible high-energy upgrade, as well as a future circular collider. For $Z'$ couplings compatible with current dijet constraints the multilepton signals can reach the $5\sigma$ level already at Run 2 of the LHC. At future colliders, couplings two orders of magnitude smaller than the electroweak coupling can be probed with $5\sigma$ sensitivity.
\end{abstract}

\section{Introduction}

One of the most attractive and popular frameworks for dark matter (DM) is the so-called $Z'$-portal \cite{Langacker:1984dc,Langacker:2008yv,FileviezPerez:2010gw,Frandsen:2011cg,Alves:2013tqa,Arcadi:2013qia,Lebedev:2014bba,Duerr:2014wra,Kahlhoefer:2015bea,Duerr:2015vna,Okada:2016gsh,Jacques:2016dqz,Fairbairn:2016iuf,Arcadi:2017hfi,Blanco:2019hah}, in which the DM particle, typically a fermion $\chi$, singlet under the standard model (SM) gauge group, interacts with SM matter through the common interaction with a massive $Z'$ boson associated to an extra gauge group, $U(1)_{Y'}$. 
Usually, the most stringent bounds on this scenario arise from di-lepton
production at the LHC \cite{Aaboud:2017buh,Sirunyan:2018exx} and DM direct-detection (DD)
experiments \cite{Aprile:2018dbl}. This has led to consider leptophobic models, in which the only coupling of $Z'$ in the SM sector is to quarks. Likewise, spin-independent DD cross-section is dramatically 
suppressed if the $Z'$ coupling to the DM particle and/or to
the quarks is axial \cite{Lebedev:2014bba,Hooper:2014fda,Kahlhoefer:2015bea,Ismail:2016tod,Ellis:2018xal,Bagnaschi:2019djj}.
Although most of the analyses of these models have been done in the context of simplified dark matter models (SDMM), in which the DM particle, $\chi$, and the mediator, $Z'$, are the only extra fields (see e.g. \cite{Buchmueller:2014yoa}), it has been recently stressed \cite{FileviezPerez:2010gw,Duerr:2015vna,Ellis:2018xal,Caron:2018yzp,Casas:2019edt} that the ultraviolet (UV) completion of the model requires the presence of additional fields in the dark sector. Such UV completion is enormously simpler and more natural if the axial coupling of the $Z'$ boson is to the DM particle, not to the quarks \cite{Casas:2019edt}. Actually, this is the only possibility if the Higgs sector contains less than three Higgs doublets. Then, leptophobia imposes that the $U(1)_{Y'}$ charge of the quarks must be universal, which means that this symmetry is identical to  baryon number in the SM sector. Concerning the dark sector, besides the DM particle, i.e. the SM singlet $\chi$, the minimal set of additional particles required to cancel all the anomalies consists of a SU(2) doublet, $\psi$, and a SU(2) singlet, $\eta$, both with non-vanishing $U(1)_Y$ and $U(1)_{Y'}$ charges~\cite{FileviezPerez:2010gw,Ellis:2017tkh,Caron:2018yzp}. Moreover, there must be at least one extra scalar, $S$, whose vacuum expectation value (VEV) breaks the $U(1)_{Y'}$ group.

There are many possible assignments of the extra hypercharges in the dark sector consistent with anomaly cancellation, but only a few leading to axial $Z'$ coupling of $\chi$ \cite{Duerr:2014wra,Ellis:2017tkh,Caron:2018yzp}. Among them, there is essentially only one in which a unique scalar gives mass not only to the $Z'$ boson, but also to all particles in the dark sector, and avoids the presence of electrically charged stable particles~\cite{Caron:2018yzp}.

The goal of this paper is to study the phenomenology of this model and explore its most distinctive signals at the LHC and future colliders. A preliminary analysis was performed in ref.~\cite{Caron:2018yzp}, assuming that the extra dark particles, $\psi$, $\eta$, were very massive, so that they decouple, leading to an effective SDMM (with a fixed correlation between the $Z'$ couplings to DM and SM).\footnote{For related work see \cite{ElHedri:2018cdm,FileviezPerez:2018jmr,FileviezPerez:2019jju}.} However, since both the DM particle, $\chi$, and the dark `leptons', $\psi$, $\eta$, obtain their masses from the same $S$-VEV, it is natural to assume that these are of the same order. Actually, this is good news, as the obliged presence of the dark leptons  offers a fortunate opportunity to test the scenario at the LHC through new and specific signals. As we will see, although the associated phenomenology has some similarities with that of supersymetric models, it also presents drastic differences, which motivate novel analyses of experimental beyond-the-SM signals.

As a matter of fact, the presence of the extra leptons not only affects the LHC phenomenology but may also modify  the production of DM in the early universe. This happens in particular if the masses of any of these extra particles is close enough to the DM one to produce non-negligible co-annihilation effects. This enhances the region of the parameter space consistent with the DM relic density and, as we will see, improves the chances to detect the scenario at the LHC.

In section~\ref{sec:2} we write down the model and the interactions of the dark leptons. The constraints on the model parameters from dark matter relic density and direct detection are examined in section~\ref{sec:3}. With these constraints in mind, we address in section~\ref{sec:4} the general features of the four-lepton signal we are interested in. Representative benchmark points are chosen in section~\ref{sec:5}, for which a detailed simulation is performed in section~\ref{sec:6}. The discussion of our results and possible implications for experimental searches are given in section~\ref{sec:7}.

\section{The model}
\label{sec:2}

\subsection{Matter content, Lagrangian and Spectrum}

The simplest extension of the SM that accommodates a leptophobic $Z'$ with axial coupling to DM has the following characteristics. The extra $U(1)_{Y'}$ gauge group is equivalent to baryon number in the SM sector (required by leptophobia). Regarding the dark sector, this consists of three (Dirac) fermions, $\chi, \psi, \eta$, with
the following $SU(2)_L\times U(1)_Y\times U(1)_{Y'}$ representations:
\begin{align}
& \chi_L\ (\,1,\,0,\,\frac{9}{2}Y'_q\,) \,
&&\chi_R\ (\,1,\,0,\, -\frac{9}{2}Y'_q\,) \,, \nonumber\\
& \psi_L\ (\,2,\, -\frac{1}{2},\,-\frac{9}{2}Y'_q\,) \,,
&&\psi_R\ (\,2,\,-\frac{1}{2},\, \frac{9}{2}Y'_q\,) \,, \nonumber\\
& \eta_L\ (\,1,\, -1,\,\frac{9}{2}Y'_q\,) \,,
&& \eta_R\ (\,1,\, -1,\, -\frac{9}{2}Y'_q\,) \,,
\label{bench}
\end{align}
where $Y'_q$ is the extra-hypercharge of the quarks, assumed positive, which if desired can be taken with the same normalisation as baryon number, i.e. $Y'_q=1/3$. All the previous fields are colour singlets, while in the SM sector only the quarks have non-vanishing $Y'$ hypercharge. Notice that all the above fields, except $\chi$, present SM-gauge interactions, so they are not `dark' in a strict sense. As we see below, there is an accidental $Z_2$ symmetry (actually a `dark leptonic number') which prevents these fermions from decaying into SM ones.
The specific charge-assignment (\ref{bench}) was first explicitly considered in \cite{Duerr:2015vna}.
The state $\chi$ is the one to naturally play the role of DM.
Flipping the signs of the above ordinary hypercharges, i.e. $Y_\psi\rightarrow -Y_\psi$, $Y_\eta\rightarrow -Y_\eta$ (independently) also  leads to a consistent model, so there are in fact four minimal models with very similar characteristics; and we will focus in the one defined by the previous assignments. In addition, the scalar sector must contain a singlet $S$, whose VEV breaks the $U(1)_{Y'}$ group.\footnote{There might exist extra scalar states, but for the study of dark lepton signals from $Z'$ boson decays performed in this paper it is enough to work in the simplest case with just one complex scalar singlet, $S$. For a discussion of $Z'$ cascade decays into scalars in a model with two singlets see ref.~\cite{Aguilar-Saavedra:2019adu}. }
Requiring that the same VEV provides masses to the dark particles fixes the charges of $S$,
\bea
S\ (\,1,\,0,\, -9Y'_q\,) \,.
\eea
Let us discuss now the most relevant pieces of the Lagrangian. The Yukawa-like terms involving the dark fermions read
\begin{eqnarray}
\mathcal{L}_Y & = & - y_1 \bar \psi_L  \eta_R \phi - y_2 \bar \psi_L \chi_R  \tilde \phi - y_3 \bar \eta_L  \psi_R \phi^\dagger - y_4 \bar \chi_L \psi_R \tilde \phi^\dagger  \notag \\
& & - \lambda_\psi \bar \psi_L \psi_R S - \lambda_\eta \bar \eta_L \eta_R S^* - \lambda_\chi \bar \chi_L \chi_R S^* + \mathrm{H.c.} \,,
\label{ec:LY}
\end{eqnarray}
with $\phi$ the SM Higgs doublet, and $\tilde \phi = i \sigma_2 \phi^*$ in standard notation.
We have assumed that the couplings $y_i$, $\lambda_i$ are real. Note that the previous Lagrangian presents an obvious, accidental $Z_2$ symmetry involving the fermionic fields.
Let us mention that there are two additional terms, consistent with the gauge symmetry, that could be added to the previous Lagrangian, namely $ - \lambda_L \bar \chi_L \chi_L S - \lambda_R \bar \chi_R \chi_R S^*$ plus their Hermitian conjugate. These terms induce a splitting of the two lightest degrees of freedom of the DM particle, thus spoiling its axial coupling to the $Z'$. Fortunately, they can be safely avoided by noticing that their absence is protected by a global `dark lepton number' under which all dark fermions, $\chi, \psi, \eta$, transform with the same charge. Let us also note that the mixing terms in the first line of (\ref{ec:LY}) are crucial to enable the decay of the electrically-charged dark fermions, which otherwise would lead to cosmological disasters.\footnote{
Without this requirement there is another consistent assignment of ordinary hypercharges in eq.~(\ref{bench}), namely $Y_\psi=\pm 7/2, Y_\eta=\pm 5$ \cite{Caron:2018yzp}. Hence, the viability of this alternative model requires extra Higgs states in order to present analogous mixing terms.}

The relevant terms of the scalar Lagrangian involving the $S-$field read
\bea
{\cal L}_{\rm scal} &\supset&  - m_S^2|S|^2 -\lambda_S^2|S|^4 - \lambda_{HS}^2|H|^2|S|^2 .
\label{Lscal}
\eea
The mixing term is constrained by Higgs measurements~\cite{ATLAS:2019slw} and does not play any relevant role in this analysis. The other two parameters can be traded by the $S$ mass and VEV.

Let us now examine the spectrum of the model after symmetry breaking. When the scalars acquire a VEV,
\begin{equation}
\phi \to \frac{1}{\sqrt 2} \left(\! \begin{array}{c} 0 \\ v \end{array} \! \right) \,,\quad S \to \frac{v_s}{\sqrt 2} \,,
\end{equation}
%
%
%====
%
%
%\section{Constraints from Dark matter. Benchmarks}
%
%
the mass terms for the dark leptons are
\begin{align}
\mathcal{L} = &
- (\bar \chi_L \, \bar \psi_L^0 ) \; \frac{1}{\sqrt 2}
\left( \! \begin{array}{cc} \lambda_\chi v_s & y_4 v \\ y_2 v & \lambda_\psi v_s \end{array} \! \right)
\left(\! \begin{array}{c} \chi_R \\ \psi_R^0  \end{array} \! \right)
 - (\bar \eta_L \, \bar \psi_L^- ) \; \frac{1}{\sqrt 2}
\left( \! \begin{array}{cc} \lambda_\eta v_s & y_3 v \\ y_1 v & \lambda_\psi v_s \end{array} \! \right)
\left(\! \begin{array}{c} \eta_R \\ \psi_R^-  \end{array} \! \right)
 + \mathrm{H.c.}
 \label{ec:LM}
\end{align}
We label the neutral mass eigenstates as $N_{1,2}$ and the charged ones as $E_{1,2}$, with masses $m_{N_1} \leq m_{N_2}$ and $m_{E_1} \leq m_{E_2}$. The relation with weak eigenstates is
\begin{align}
& \left( \! \begin{array}{c} N_{1L,1R} \\ N_{2L,2R} \end{array} \! \right) =  U_{L,R}^N 
\left( \! \begin{array}{c} \chi_{L,R} \\ \psi_{L,R}^0 \end{array} \! \right) =
\left( \! \begin{array}{cc} \cos \theta_{L,R}^N & -\sin \theta_{L,R}^N \\ \sin \theta_{L,R}^N & \cos \theta_{L,R}^N \end{array} \! \right)
\left( \! \begin{array}{c} \chi_{L,R} \\ \psi_{L,R}^0 \end{array} \! \right) \,,  \notag \\[1mm]
& \left( \! \begin{array}{c} E_{1L,1R} \\ E_{2L,2R} \end{array} \! \right) =  U_{L,R}^E 
\left( \! \begin{array}{c} \eta_{L,R} \\ \psi_{L,R}^- \end{array} \! \right) =
\left( \! \begin{array}{cc} \cos \theta_{L,R}^E & -\sin \theta_{L,R}^E \\ \sin \theta_{L,R}^E & \cos \theta_{L,R}^E \end{array} \! \right)
\left( \! \begin{array}{c} \eta_{L,R} \\ \psi_{L,R}^- \end{array} \! \right) \,.
\label{ec:Udef}
\end{align}
The lightest neutral eigenstate $N_1$ is the dark matter candidate. Defining $r=v/v_s$, the mixing angles for the neutral sector are given by
\begin{equation}
\tan 2 \theta_L^N = 2 r \frac{\lambda_\chi y_2 + \lambda_\psi y_4}{\lambda_\psi^2 - \lambda_\chi^2 + r^2 (y_2^2 - y_4^2)} \,,\quad
\tan 2 \theta_R^N = 2 r \frac{\lambda_\chi y_4 + \lambda_\psi y_2}{\lambda_\psi^2 - \lambda_\chi^2 - r^2 (y_2^2 - y_4^2)} \,.
\label{ec:thdef}
\end{equation}
The mixing angles in the charged sector $\theta_{L,R}^E$ have analogous expressions with the replacements $y_2 \to y_1$, $y_4 \to y_3$, $\lambda_\chi \to \lambda_\eta$.
The four masses $m_{N_{1,2}}$, $m_{E_{1,2}}$ and four mixing angles $\theta_{L,R}^{N,E}$ are not independent parameters, and satisfy the relation
\begin{equation}
m_{E_1} \sin \theta_L^E \sin \theta_R^E + m_{E_2} \cos \theta_L^E \cos \theta_R^E =
m_{N_1} \sin \theta_L^N \sin \theta_R^N + m_{N_2} \cos \theta_L^N \cos \theta_R^N
\end{equation}
that stems from the equality of the $(2,2)$ entries of the neutral and charged lepton mass matrices of eq.~(\ref{ec:LM}).

Typically the mixing angles are small (or close to $\pi/2$) if the Yukawa couplings ($y_2, y_4$ for $\theta_{L,R}^N$; $y_3, y_3$ for $\theta_{L,R}^E$) are small. As we will see, from DM direct-detection bounds, section~\ref{sec:3}, this is indeed the expected situation for the neutral angles, $\theta_{L,R}^N$. Consequently, we expect the DM particle, $N_1$, to be mostly $\chi$-like.

\vspace{0.3cm}
The expressions for the mass eigenvalues are lengthy and not very illuminating, but they get greatly simplified in the limit where left and right angles are equal, which occurs for 
\begin{align}
& y_2 = y_4 \quad \Rightarrow \quad \theta_L^N = \theta_R^N \equiv \theta^N \,, \notag \\
& y_1 = y_3 \quad \Rightarrow \quad \theta_L^E = \theta_R^E \equiv \theta^E \,.
\end{align}
Actually, this assumption has very mild implications on the collider phenomenology, the most important effect being the modification of angular distributions in decay chains.
%, besides the fact that, as we will see in the next section, in the neutral sector $y_2 = y_4$ implies very small mixing angles. The latter however do not render any of the heavier particles $N_2$, $E_1$, $E_2$ long-lived so as to produced displaced vertices. 
With this simplification, one can obtain compact exact expressions for the masses,
\begin{align}
& m_{N_1} =m_\chi  - \Delta_N \,,
&& m_{E_1} = m_\psi  - \Delta_E \,,\notag \\
& m_{N_2} = m_\psi  + \Delta_N \,,
&& m_{E_2} = m_\eta + \Delta_E \,,
\label{M1}
\end{align}
with
\be
m_\chi = \frac{1}{\sqrt{2}}\ \lambda_\chi v_s,\ \   m_\psi = \frac{1}{\sqrt{2}}\  \lambda_\psi v_s,\ \   m_\eta = \frac{1}{\sqrt{2}}\  \lambda_\eta v_s \,,
\label{M2}
\ee
and
\begin{align}
& \Delta_N = y_2 \frac{v}{\sqrt 2} \tan \theta^N
= (m_\psi - m_\chi) \frac{\sin^2 \theta^N}{\cos 2 \theta^N} \,, \notag \\
& \Delta_E = y_1 \frac{v}{\sqrt 2} \tan \left(\frac{\pi}{2} - \theta^E \right) 
= (m_\psi- m_\eta)  \frac{\cos^2 \theta^E}{\cos 2 \theta^E} \,.
\label{M3}
\end{align}

\subsection{Interactions in the mass basis}
%\label{sec:2}

The interactions of the dark leptons with the various gauge bosons in the weak basis, $\left\{\chi, \psi, \eta\right\}$,
 are either vectorial or axial, see eq.~(\ref{bench}). In the mass eigenstate basis, $\left\{N_{1,2}, E_{1,2}\right\}$, they remain with this character provided the left- and right-handed mixing angles are equal.\footnote{This includes the case where both are very small. As we will see in section~\ref{sec:3}, this is a very reasonable limit, especially for the neutral angles. Hence the interaction of the DM  with the $Z'$ boson is expected to maintain its axial character.}
In general, the interactions of dark leptons with the $Z'$ boson can be written as
\begin{eqnarray}
\mathcal{L}_{Z'} & = &
 - g_{Z'}  Y^\prime_{F}  (  \mathcal{Z}_{ij}^{NL} \bar N_{iL}  \gamma^\mu N_{jL} 
 -  \mathcal{Z}_{ij}^{NR} \bar N_{iR}  \gamma^\mu N_{jR} 
 +  \mathcal{Z}_{ij}^{EL} \bar E_{iL}  \gamma^\mu E_{jL}  \notag \\
& & -  \mathcal{Z}_{ij}^{ER} \bar E_{iR}  \gamma^\mu E_{jR}  ) B'_\mu  \,,
\end{eqnarray}
with $i,j=1,2$, $F=E,N$, $Y'_F = 9/2 \, Y'_q$. The mixing parameters for the left-handed neutral leptons are given by
\begin{align}
& \mathcal{Z}_{11}^{NL} = \cos 2 \theta_L^N \,,  \notag \\
&  \mathcal{Z}_{22}^{NL} = -  \cos 2 \theta_L^N \,, \notag \\
&  \mathcal{Z}_{12}^{NL} = \mathcal{Z}_{21}^{NL} = \sin 2 \theta_L^N \,.
\end{align}
For the right-handed sector they have the same expressions but replacing $\theta_L^N$ by the corresponding angle $\theta_R^N$. The mixing parameters for charged fields can be obtained simply by replacing the neutral mixing angles $\theta_{L,R}^N$ by $\theta_{L,R}^E$. The interactions with the $W$ boson read
\begin{eqnarray}
\mathcal{L}_{W} & = &
 - \frac{g}{\sqrt 2}  (\mathcal{V}_{ij}^L \bar N_{iL} \gamma^\mu E_{jL} + \mathcal{V}_{ij}^R \bar N_{iR} \gamma^\mu E_{jR} ) W_\mu^+ + \text{H.c.} 
\end{eqnarray}
The left-handed mixing parameters are
\begin{align}
& \mathcal{V}_{11}^{L} = \sin \theta_L^N \sin \theta_L^E \,, \notag \\
& \mathcal{V}_{22}^{L} = \cos \theta_L^N \cos \theta_L^E \,, \notag \\
& \mathcal{V}_{12}^{L} =  - \sin \theta_L^N \cos \theta_L^E \,, \notag \\
& \mathcal{V}_{21}^{L} = - \cos \theta_L^N \sin \theta_L^E \,.
\end{align}
The expressions for right-handed mixings $\mathcal{V}_{ij}^R$ are the same as for $\mathcal{V}_{ij}^L$ above but replacing $\theta_L^{N,R}$ by $\theta_R^{N,E}$. 
The interactions with the $Z$ boson read
\begin{eqnarray}
\mathcal{L}_{Z} & = &
 - \frac{g}{2 c_W} (\mathcal{X}_{ij}^{NL} \bar N_{iL}   \gamma^\mu N_{jL} + \mathcal{X}_{ij}^{NR} \bar N_{iR}   \gamma^\mu N_{jR} ) Z_\mu  \notag \\
& &  + \frac{g}{2 c_W} (\mathcal{X}_{ij}^{EL}  \bar E_{iL}  \gamma^\mu E_{jL}  + \mathcal{X}_{ij}^{ER}  \bar E_{iR}  \gamma^\mu E_{jR} ) Z_\mu  \,,
\end{eqnarray}
where the left-handed mixing parameters are
\begin{align}
& \mathcal{X}_{11}^{NL} = \sin^2 \theta_L^N \,, 
&& \mathcal{X}_{11}^{EL} = \sin^2 \theta_L^E - 2 s_W^2 \,, \notag \\
&  \mathcal{X}_{22}^{NL} =  \cos^2 \theta_L^N \,, 
&& \mathcal{X}_{22}^{EL} =  \cos^2 \theta_L^E - 2 s_W^2 \,, \notag \\
&  \mathcal{X}_{12}^{NL} = \mathcal{X}_{21}^{NL} = - \oh \sin 2 \theta_L^N \,, 
&& \mathcal{X}_{12}^{EL} = \mathcal{X}_{21}^{EL} = - \oh \sin 2 \theta_L^E \,,
\end{align}
and the right-handed counterparts have similar expressions but replacing $\theta_L^{N,E}$ by $\theta_R^{N,E}$. 
Photon interactions are flavour-diagonal,
\begin{equation}
\mathcal{L}_\gamma = e ( \bar E_1 \gamma^\mu E_1 + \bar E_2 \gamma^\mu E_2 ) A_\mu \,,
\end{equation}
The interactions with the Higgs boson arise from the terms in the first line of (\ref{ec:LY}). In the mass eigenstate basis,
\begin{equation}
\mathcal{L}_H = - \left[  \mathcal{Y}_{ij}^N \bar N_{iL} N_{jR} +  \mathcal{Y}_{ij}^E \bar E_{iL} E_{jR} \right] H + \text{H.c.}
\label{FFH}
\end{equation}
For convenience, the Yukawa couplings $\mathcal{Y}_{ij}^{N,E}$ can be parameterised in terms of masses and mixing angles. For the neutral sector they are
\begin{align}
& \mathcal{Y}_{11}^N = \frac{m_{N_1}}{2v} [ 1 - \cos 2\theta_L^N \cos 2 \theta_R^N] -   \frac{m_{N_2}}{2v} \sin 2 \theta_L^N \sin 2 \theta_R^N \,, \notag \\
&  \mathcal{Y}_{22}^N =  \frac{m_{N_2}}{2v} [ 1 - \cos 2\theta_L^N \cos 2 \theta_R^N] -   \frac{m_{N_1}}{2v} \sin 2 \theta_L^N \sin 2 \theta_R^N \,, \notag \\
&  \mathcal{Y}_{12}^N = - \frac{m_{N_1}}{2v} \cos 2 \theta_L^N \sin 2 \theta_R^N +  \frac{m_{N_2}}{2v} \sin 2 \theta_L^N \cos 2 \theta_R^N \,, \notag \\
&  \mathcal{Y}_{21}^N = - \frac{m_{N_1}}{2v} \sin 2 \theta_L^N \cos 2 \theta_R^N +  \frac{m_{N_2}}{2v} \cos 2 \theta_L^N \sin 2 \theta_R^N \,.
\label{FFH2}
\end{align}
For the charged leptons the Yukawa couplings $\mathcal{Y}_{ij}^E$ have similar expressions but replacing the masses and mixing angles by the corresponding ones in the charged sector.

\section{Constraints from Dark matter}
\label{sec:3}

The thermal relic abundance of DM is determined by the efficiency of the processes that  lead to its annihilation in the early universe. In the first place, there are the processes mediated by the $Z'$ boson, in particular $\chi \bar \chi \rightarrow Z'\rightarrow q\bar q$ (recall here that the DM particle, $N_1$, is close to a pure $\chi$ state). Besides, for heavy enough DM there are processes $\chi \bar \chi \rightarrow Z' Z'$ with a $\chi$ in $t-$channel. The last case, however, does not apply to the instances examined in this paper, where the dark matter is much lighter than the $Z'$ boson. More precisely, as discussed in section~\ref{sec:5}, we will consider DM masses and splittings between masses of dark leptons in the few-hundred GeV range, while $m_{Z'}$ and $m_S$ will be in the few TeV range.

In ref.\cite{Caron:2018yzp} it was shown that, in order to reproduce the observed relic abundance, the  $g_{Z'}$ coupling involved in the previous processes must be fairly sizeable. This in turn leads to strong experimental bounds coming from di-jet production at LHC. Actually, there is a broad range of $Z'$ masses, $500\ {\rm GeV}\simlt m_{Z'}\simlt 3500\ {\rm GeV}$, which is excluded on these grounds. If the scalar associated to the $S$ field is sufficiently light, there are additional annihilation processes in play, which slightly reduces the required value of $g_{Z'}$, leading to a (modest) enhancement of the allowed region. This situation makes challenging to probe the scenario at the LHC, since the resonant production of the new particles occurs essentially beyond the present energy limit. 

On the other hand, the presence of the extra fields, $\psi, \eta$ (or, more precisely, $N_2$, $E_1$, $E_2$) offers new possibilities to annihilate DM in the early universe, something not considered in ref. \cite{Caron:2018yzp}. The most obvious one is the co-annihilation of the DM particle with one of these states. In this sense, the most convenient state to play this role is $\psi$, not only for the possibility of direct co-annihilations, but also because the direct interactions between $\chi$ and $\psi$ in the Lagrangian (\ref{ec:LY}) keep naturally the DM in thermal equilibrium with these extra degrees of freedom.\footnote{We are referring here to co-annihilation in a generic sense, which includes not only co-annihilation stricto sensu, but also the transfer from the $\chi$ population to the  $\psi$ one (thanks to the thermal equilibrium), which is subsequently annihilated through much more efficient (weak-interaction) processes, see refs.~\cite{DAgnolo:2017dbv,Garny:2017rxs}.} This additional source of annihilation relaxes the required value of $g_{Z'}$ in order to get the correct relic density. Actually, for $m_\psi$ close enough to $m_\chi$ there is no even need of the $Z'$-mediated contribution to the annihilation. This means that the value of $g_{Z'}$ becomes in practice a free parameter, provided the gap between the two masses is the suitable one to produce the necessary amount of co-annihilation.

Co-annihilation processes are very sensitive to the mass gap between the DM particle and the co-annihilating one. For example, in our case, for $m_{Z'} = 2.5$ TeV, $m_\chi=300$ GeV, $m_\psi=313.9$ GeV, the observed relic density is entirely obtained thanks to co-annihilation processes, thus $g_{Z'}$ must be rather small to avoid an excess of annihilation. 
Decreasing $m_\psi$ further makes the co-annihilation too efficient, so that the relic density falls below the observed value. On the other hand, increasing $m_\psi$, the efficiency of the co-annihilation drops quickly and the relic density becomes too large. This can be fixed by an appropriate increase of $g_{Z'}$, and thus of the efficiency of the annihilation processes mediated by $Z'$. However, increasing $m_\psi$ in just 0.1 GeV requires to raise $g_{Z'}$ above perturbative levels. For LHC phenomenology this means that it is enough to set the value of $m_\psi$ at this narrow range and leave $g_{Z'}$ as a free-parameter. Notice also that the required value of $m_\psi$ is essentially independent of $m_{Z'}$, since the annihilation of the $\psi$ states mainly involve  weak interactions.

A scenario of co-annihilation as the one depicted above requires a mass-ordering $m_\chi<m_\psi<m_\eta$ , which implies in turn that $|\theta_{L,R}^N|\in [0, \pi/4]$, $\theta_{L,R}^E\in [\pi/4, 3\pi/4]$, where, for convenience, we have taken the definition ranges of the angles as $-\pi/2 \leq \theta_{L,R}^N\leq \pi/2$, $0 \leq \theta_{L,R}^E\leq \pi$.
In the limit $y_i\rightarrow 0$ the angles become  $\theta_{L,R}^N=0$, $\theta_{L,R}^E=\pi/2$. Note that the reason for the latter is simply that 
$m_\psi<m_\eta$ in the charged mass matrix (\ref{ec:LM}), while by definition $m_{E_1} \leq m_{E_2}$. 
 
Concerning the constraints from direct detection, the axial (vectorial) coupling of the  $Z'$ mediator to the DM particle (the quarks) leads to spin-dependent, velocity-suppressed DM--nucleon cross section, which is safe from present DD experimental bounds. However, the mixing of $\chi$ and $\psi$ inside the DM particle, $N_1$, leads to a non-vanishing $N_{1L} N_{1R} H$ coupling, which is dangerous since it induces spin-independent cross section. The size of this coupling can be read from eqs.~(\ref{FFH}), (\ref{FFH2}).
The corresponding bounds on $y_2, y_4$ from DD exclusion limits are very strong. In particular, for $y_2=y_4\equiv y$ (the case in which $\theta_L^N = \theta_R^N$), the bound for the previous example is $y^2 \leq 4 \times 10^{-6}$. Consequently, as mentioned in previous sections, from DD constraints one expects very small angles, $|\theta_{L,R}^N| \lesssim 0.05$, see  eq.~(\ref{ec:thdef}). In contrast, the size of the $\theta_L^E$, $\theta_R^E$ angles is no restricted by DM phenomenology.
There are additional one-loop-induced electroweak processes that contribute to DD by the interchange of a $Z-$boson. However, beside the $(4\pi)^{-2}$ suppression, those processes involve two $y-$couplings (since $\chi$ does not have direct EW interactions), and are thus negligible.

Concerning indirect detection constraints, the  most  stringent limits are currently given by a recent combined analysis of imaging air Cherenkov telescope (IACT) arrays: HESS, MAGIC, VERITAS; the Fermi-LAT satellite, and the water Cherenkov detector HAWC of the $\gamma$-ray emission in dwarf spheroidal galaxies \cite{Oakes:2019ywx},  which  can  provide  very  strong  limits on the DM annihilation cross-section $\left < \sigma v \right >$. As it has been stated above the relevant DM annihilation channel for this model is to a pair of quarks mediated by a $Z'$. Since the annihilation cross-section for this process at present time is strongly velocity-suppressed~\cite{Arcadi:2014lta},  the limits from dwarf galaxies are not effective in this model.

\section{Features of the four-lepton signal}
\label{sec:4}

The fermionic decay modes of the $Z'$ have partial widths
\begin{eqnarray}
\Gamma(Z' \to q \bar q) & = & \frac{N_c (g_{Z'} Y'_q)^2}{12 \pi} M_{Z'}
\left[ 1+2 \frac{m_q^2}{M_{Z'}^2} \right] \left[1-4 \frac{m_q^2}{M_{Z'}^2} \right]^{1/2} \,, \notag \\
\Gamma(Z' \to F_i F_j) & = & \frac{(g_{Z'} Y_F^\prime )^2}{24 \pi M_{Z'}} \lambda^{1/2}(M_{Z'}^2,m_{F_i}^2,m_{F_j}^2) \left\{ [ (\mathcal{Z}_{ij}^{FL})^2 + (\mathcal{Z}_{ij}^{FR})^2 ]
\left[ 1 - \frac{m_{F_i}^2 + m_{F_j}^2}{M_{Z'}^2}  \right. \right. \notag \\
&&  \left. \left. -  \frac{m_{F_i}^4 + m_{F_j}^4}{M_{Z'}^4} + \frac{m_{F_i}^2 m_{F_j}^2}{M_{Z'}^4} \right]
+ 6  \mathcal{Z}_{ij}^{FL} \mathcal{Z}_{ij}^{FR} \frac{m_{F_i} m_{F_j}}{M_{Z'}^2} \right\}
 \,.
\end{eqnarray}
If the mixing angles in the left and right-handed sector are equal, then $\mathcal{Z}_{ij}^{FL} = \mathcal{Z}_{ij}^{FR} \equiv \mathcal{Z}_{ij}^{F}$ and the latter equation simplifies to
\begin{eqnarray}
\Gamma(Z' \to F_i F_j) & = & \frac{g_{Z'}^2 Y_{F}^{\prime \, 2}}{12 \pi M_{Z'}} \lambda^{1/2}(M_{Z'}^2,M_{F_i}^2,M_{F_j}^2)  (\mathcal{Z}_{ij}^{F})^2
\left[ 1 - \frac{m_{F_i}^2 + m_{F_j}^2 }{M_{Z'}^2} + 3 \frac{m_{F_i} m_{F_j}}{M_{Z'}^2}  \right. \notag \\
&&  \left. -  \frac{m_{F_i}^4 + m_{F_j}^4}{M_{Z'}^4} + \frac{m_{F_1}^2 m_{F_2}^2}{M_{Z'}^4} \right]
 \,.
\end{eqnarray}
In the limit in which the $Z'$ boson is much heavier than its decay products,
\begin{equation}
\sum_{F_i,F_j} \Gamma (Z' \to F_i F_j) = \frac{9}{2} \sum_q \Gamma(Z' \to q \bar q) \,,
\end{equation}
and the $Z'$ branching ratio to dark leptons is $9/11 \simeq 80\%$. For simplicity, we assume that the scalar singlet is heavier than $M_{Z'}/2$, so that the $Z'$ boson does not decay into scalar pairs.

\begin{figure}[htb]
\begin{center}
\begin{tabular}{ccc}
\includegraphics[height=5.2cm,clip=]{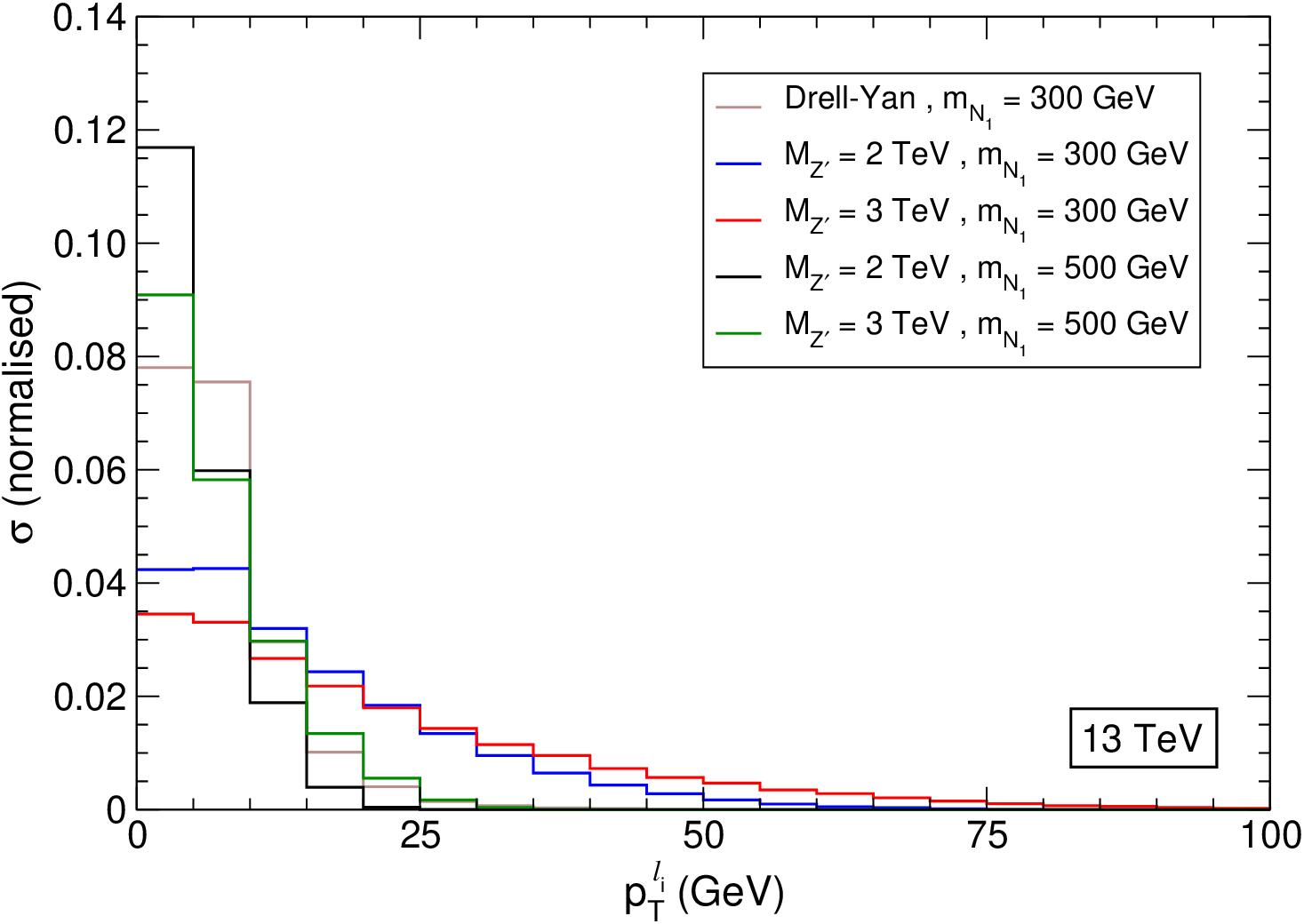} & \quad &
\includegraphics[height=5.2cm,clip=]{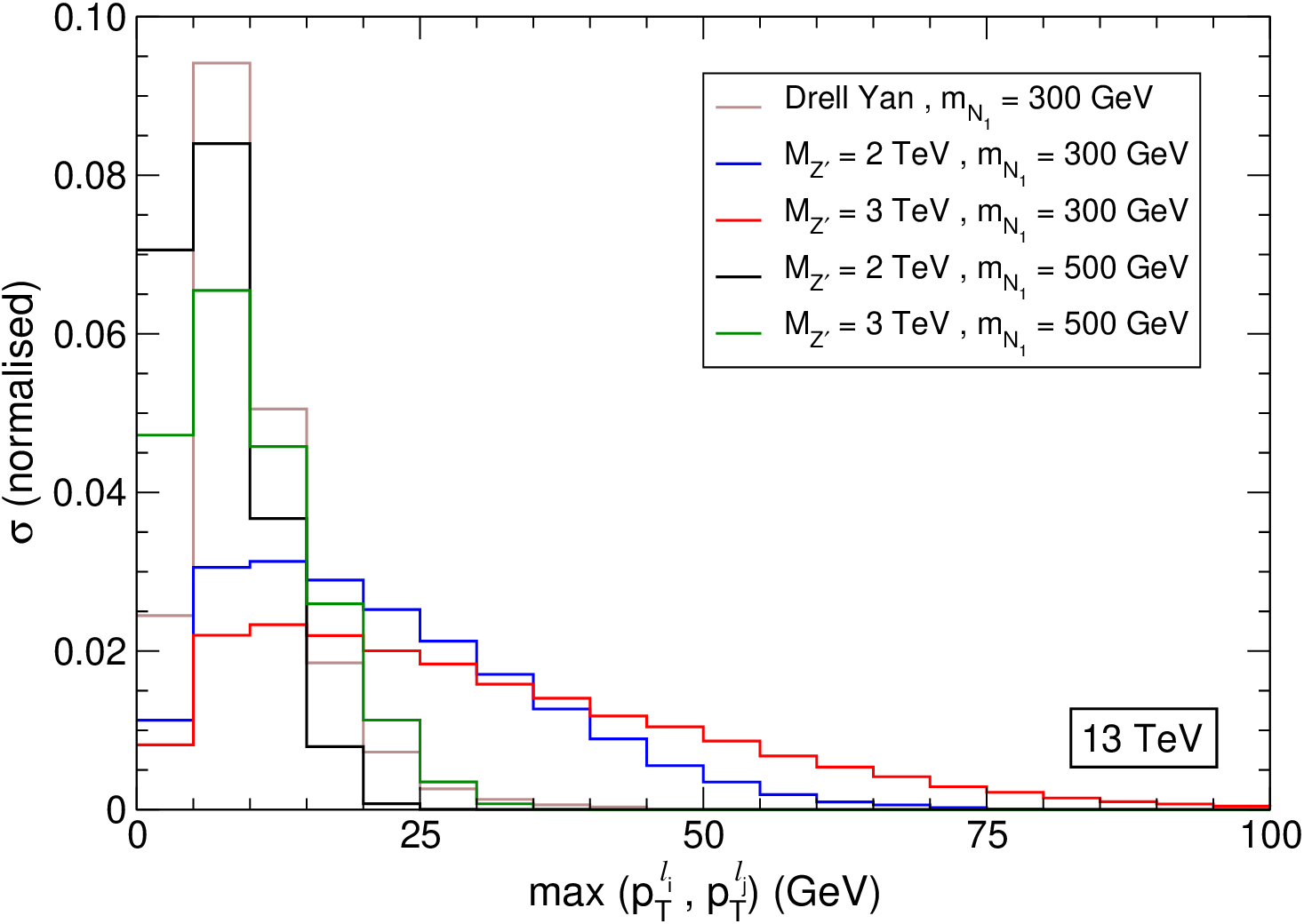} 
\end{tabular}
\caption{Left: kinematical distributions of the transverse momentum at parton level of either lepton
resulting from $pp \to Z^{(\prime)} \to N_2 N_2$, $N_2 \to N_1 \ell^+ \ell^-$. Right: kinematical distribution of the maximum of the transverse momenta of the two leptons resulting from the same $N_2$.}
\label{fig:parton1}
\end{center}
\end{figure}

\begin{figure}[htb]
\begin{center}
\includegraphics[height=5.2cm,clip=]{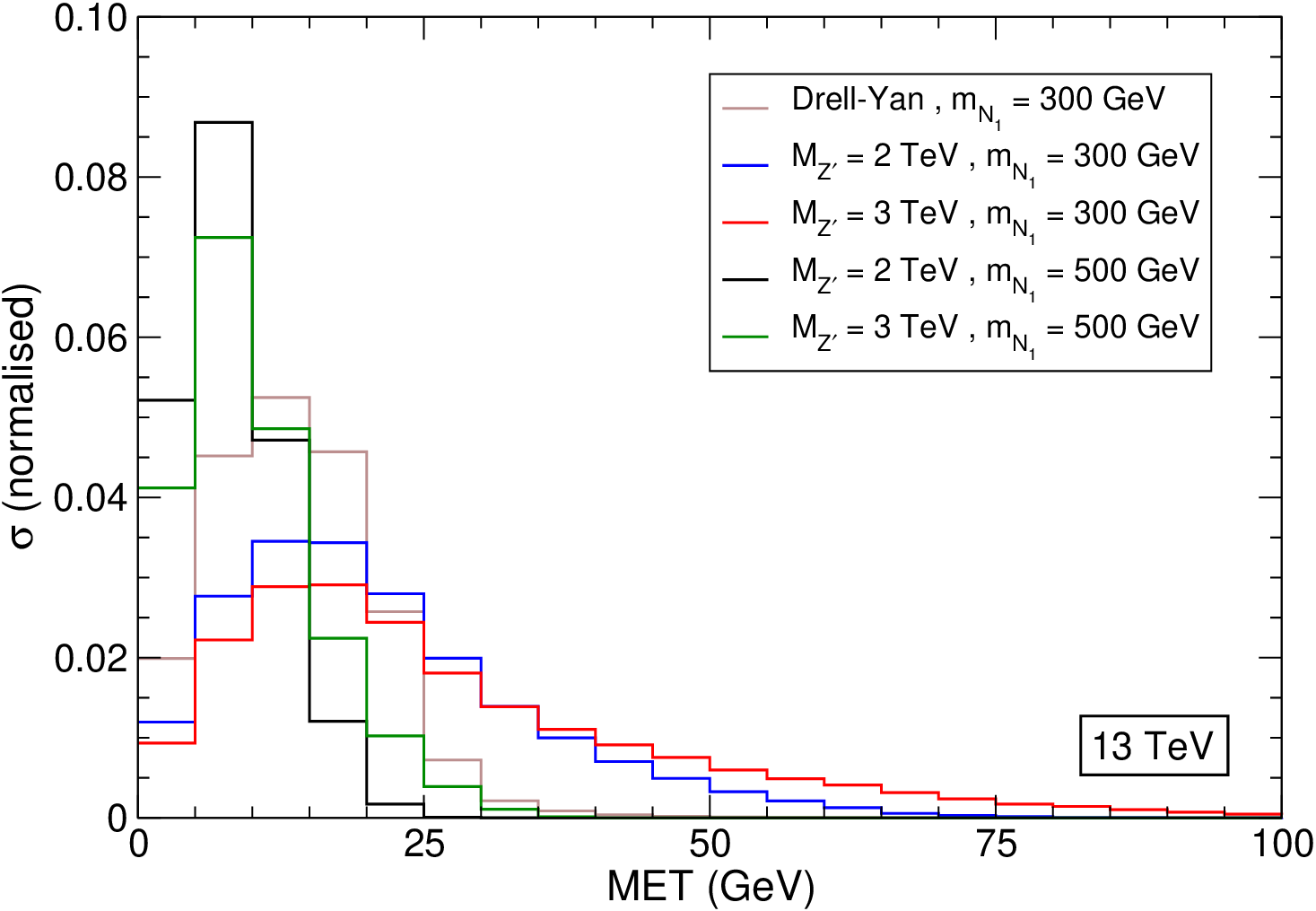} 
\caption{Kinematical distributions of the missing energy (at parton level) in $pp \to Z^{(\prime)} \to N_2 N_2$.}
\label{fig:parton2}
\end{center}
\end{figure}

The most promising signal for the kind of scenario analysed here is the production of four leptons in the final state through the process
\be
pp \rightarrow Z' \rightarrow N_2 N_2
\ee
and the subsequent leptonic decays, $N_2 \to N_1 \ell^+ \ell^-$. This final state provides the best balance between signal branching ratio and SM background.\footnote{Drell-Yan pair production modes of dark leptons have cross sections that are comparable, for the $Z'$ masses considered here, but they produce leptons with very low transverse momentum, as shown in figures~\ref{fig:parton1} and \ref{fig:parton2}. As discussed at the end of section~\ref{sec:6}, such signals are likely unobservable.}
We have also considered three-lepton signals, e.g. from $Z' \to N_2 N_2$ when one of the charged leptons does not pass the minimum $p_T$ requirement. Unfortunately, these signals are swamped by the $WZ$ background. Even worse is the situation for two-lepton signals from $Z' \to E_1 E_1$, $E_1 \to N_1 \ell \nu$. We have also investigated five-lepton signals from $Z' \to E_2 E_2 \to N_2 W \, N_2 W$, with one $W$ boson decaying hadronically and the other one leptonically. Despite the five-lepton signal is very clean, its branching ratio is too small to be competitive with the four-lepton one.

The features and visibility of this four-lepton signal essentially depend  on four parameters, the $Z'$ mass and coupling and the two neutral lepton masses, $m_{N_{1,2}}$, in a non-trivial and entangled way (the mixing angles may also affect the signal by modifying the branching ratios and angular distributions). In order to better understand the dependence, we study semi-analitically their influence in this section, previous to the simulation of selected benchmarks in section~\ref{sec:5}.

The decay $N_2 \to N_1 \ell^+ \ell^-$ produces a lepton pair of invariant mass 
\be
m_{\ell \ell} \leq m_{N_2} - m_{N_1}.
\label{mll}
\ee
If $m_{N_2} \sim m_{N_1}$, as required for the co-annihilation, the distinctive signature is a small invariant mass lepton pair. Therefore, the decay $Z' \to N_2 N_2$ produces two same-flavour opposite-sign lepton pairs of small invariant mass. Moreover, most of the energy is taken by the $N_1 N_1$ pair. Let $E_\ell^*$ be the energy of either lepton in the $N_2$ rest frame, which has a maximum
\begin{equation}
E_\ell^* \leq \frac{m_{N_2}^2 - m_{N_1}^2}{2 m_{N_2}} \,.
\end{equation}
If $m_{N_2} \sim m_{N_1}$, then $E_\ell^* / m_{N_2} \leq (m_{N_2}-m_{N_1})/m_{N_2}$, which is a small fraction, and most of the energy is kept by $N_1$.
Because the $N_1$ are produced nearly at rest in the $N_2$ rest frame, in the laboratory frame the $N_1 N_1$ pair is approximately produced back-to-back, as the $N_2 N_2$ pair is. Therefore, their contribution to the missing energy cancels to a large extent. 

Although the leptons are produced from the decay of a multi-TeV resonance, their transverse momentum is relatively small.
As aforementioned, most of the energy is taken by the $N_1 N_1$ pair. The transverse momentum of the leptons $p_T^\ell$ has an upper bound
\begin{equation}
p_T^{\ell} \leq E_\ell^* \frac{M_{Z'}}{2 m_{N_2}} \left[ 1 + \left( 1-4 \frac{m_{N_2}^2}{M_{Z'}^2} \right)^\frac{1}{2} \right] \,.
\end{equation}
If $m_{N_2} \sim m_{N_1} \ll M_{Z'}$, this simplifies to
\begin{equation}
p_T^{\ell} \leq \frac{m_{N_2} - m_{N_1}}{m_{N_2}}  M_{Z'} \,.
\end{equation}
We show in figure~\ref{fig:parton1} (left) the kinematical distribution of the transverse momentum of either (positive or negative) of the leptons resulting from $pp \to Z' \to N_2 N_2$, $N_2 \to N_1 \ell^+ \ell^-$ in the laboratory frame, for four sets of values of the $Z'$ and heavy lepton masses. In the examples with $m_{N_1} = 300$ GeV we set $m_{N_2} = 314$ GeV, and in the examples with $m_{N_1} = 500$ GeV we set $m_{N_2} = 507$ GeV. These are the values of $m_{N_2}$ that provide the correct amount of relic density along the lines discussed in section~\ref{sec:3}. For comparison, we also show the transverse momentum distribution for Drell-Yan production $pp \to Z \to N_2 N_2$ where, as expected, the leptons are very soft.
In the right panel we show the kinematical distribution of the maximum of the $p_T$ for the two leptons $\ell_1 \ell_2$ from the decay of the {\it same} $N_2$. For $Z'$ decays, in a significant fraction of the events one or two of the leptons can trigger the recording of the event (there are also two other leptons with identical distributions from the decay of the other $N_2$). However, the signal efficiency would benefit from additional low-threshold four-lepton triggers.
For illustration, we show in figure~\ref{fig:parton2} the kinematical distribution of the missing energy (MET) computed at parton level, using the sum of the three-momenta of the two stable $N_1$. As anticipated, the missing energy is relatively small.

\section{Benchmarks}
\label{sec:5}

This section is devoted to formulating benchmark points in the parameter space of the model that are consistent with all the phenomenological constraints (including those from DM) and are representative of the new phenomenology that emerges from this scenario. 

Let us consider first the mixing angles $\theta_{L,R}^N$, $\theta_{L,R}^E$. Indeed, they are naturally small ($\theta_{L,R}^N$) or close to $\pi/2$ ($\theta_{L,R}^E$) due to to suppression factor $r=v/v_S$ in the expressions given around eq.~(\ref{ec:thdef}). Still, they might show a substantial departure from those values.
Note in particular that for $\theta_{L,R}^N$ the denominator in eq.~(\ref{ec:thdef}) could be quite small since $\lambda_\chi\simeq \lambda_\psi$ in order to allow an efficient co-annihilation, typically $\lambda_\psi - \lambda_\psi = {\cal O}(10^{-2})$. However, as discussed in section~\ref{sec:3}, to avoid problems with direct detection the Yukawa couplings $y_2, y_4$ must be substantially smaller, ${\cal O}(10^{-3})$, thus rendering the neutral angles, $\theta_{L,R}^N$, very small.
On the other hand, the precise values of $y_2, y_4$ are irrelevant for most of the phenomenology, provided this bound is satisfied. We will take them so that $\theta_{L}^N=\theta_{R}^N=0.02$ (more details below). 

Concerning $\theta_{L,R}^E$, although they are naturally close to $\pi/2$, they certainly could be quite different without conflicting any experimental data. This is illustrated in figure~\ref{fig:scan}, where we have fixed all the parameters as in Benchmark 1 (\ref{ec:Benchmark1}) below, except the $y_1, y_3$ couplings, and hence $\theta_{L,R}^E$. Scanning over $y_{1,3}$ with $|y_{1,3}| \leq \text{max}(\lambda_\chi,\lambda_\psi,\lambda_\eta)$ (in this case 0.51) gives the allowed blue region in the $\theta_{L}^E-\theta_{R}^E$ plane. 
The area $\theta_{L,R}^E\sim\pi/2$ is the widest one and the departure from it is bounded, but still the possibility of sizeable mixings exists. However, such situation is inconvenient to test the model at the LHC. 
The reason is that sizeable $\cos\theta_{L,R}^E$ would lead to $m_{E_1}<m_{N_2}$, as is illustrated by the expressions (\ref{M1})--(\ref{M3}). If this mass gap is not tiny, the $N_2$ state would naturally decay as $N_2\to E_1 W^*$ (instead of $ N_1 Z^*$), thus ruining the four-lepton signal. Consequently, in our benchmarks we will choose small $y_1, y_3$, so that $\theta_{L}^E=\theta_{R}^E=\pi/2 - 0.02$.

\begin{figure}[htb]
\begin{center}
\begin{tabular}{ccc}
\includegraphics[height=5.5cm,clip=]{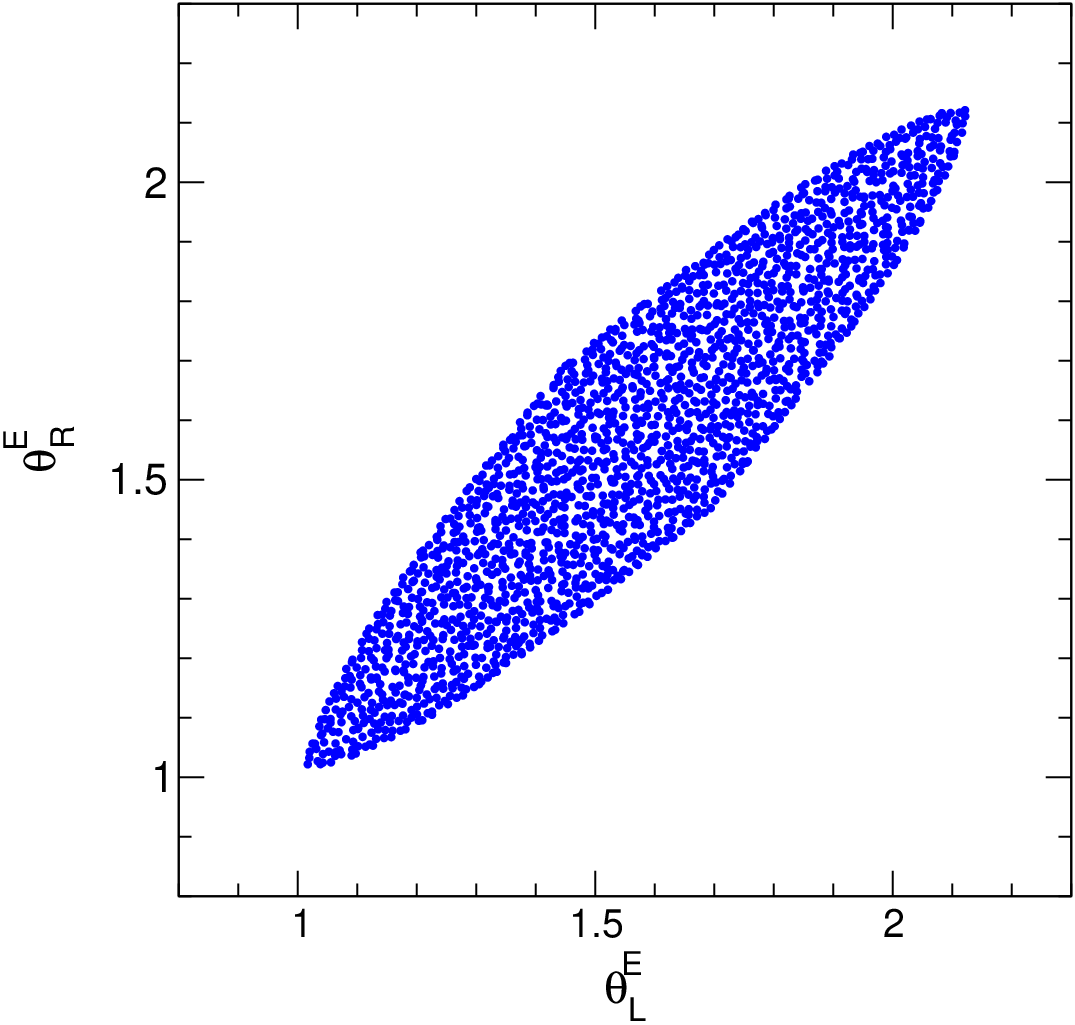} 
\end{tabular}
\caption{Allowed region for the mixing angles in the heavy charged sector for a benchmark scenario specified in eqs. (\ref{ec:Benchmark1}).}
\label{fig:scan}
\end{center}
\end{figure}

Concerning the other parameters, we will fix $g_{Z'}$ at a weak-interaction size, $g_{Z'} Y'_q=0.2$, and take two values for the mass of the extra gauge boson, namely $m_{Z'}=2, 3$ TeV. To be in the safe side we take a DM mass of 300 GeV, while the co-annihilating particle is 14 GeV heavier (obliged for a correct relic abundance). 
 We have verified with {\sc CheckMATE}~\cite{Kim:2015wza,Dercks:2016npn,Drees:2013wra} that the two benchmark points adopted for this
 study are not excluded by multi-lepton searches at LHC based on 36 fb$^{-1}$ of data collected at 
$\sqrt{s} = 13$ TeV.
Finally, the mass of the heavier dark lepton, $m_{E_2}$ has a sub-leading effect on the signals and we will fix it at 400 GeV. In summary our two benchmarks are 

\vspace{0.3cm}
\noindent 
Benchmark 1
\begin{align}
& M_{Z'} = 2~\text{TeV} \,, \quad m_{N_1} = 300~\text{GeV} \,, \quad m_{N_2} = m_{E_1} = 314~\text{GeV} \,, \quad m_{E_2} = 400~\text{GeV}\,, \notag \\
& g_{Z'} Y'_q=0.2\,, \quad  \theta_{L}^N=\theta_{R}^N=0.02\,, \quad \theta_{L}^E=\theta_{R}^E=\pi/2 - 0.02\,.
\label{ec:Benchmark1}
\end{align}

\vspace{0.3cm}
\noindent 
Benchmark 2
\begin{align}
\hspace{-5cm} M_{Z'} = 3~\text{TeV} \,,~ \text{and the same remaining parameters.}
\label{ec:Benchmark2}
\end{align}

%\vspace{0.2cm}
\noindent 
In both cases, as mentioned, we assume that the scalar $S$ is heavy enough  (i.e. $m_S>M_{Z'}/2$) to be ignored. If it were light it could be involved in additional 
decay-chains with dark leptons, a kind of signals that is out of the scope of this paper.
Of course, the previous values are obtained with appropriate choices of the parameters in the initial Lagrangian (\ref{ec:LY}). More precisely, for Benchmark 1:  
\begin{align}
& v_s = 1111~\text{GeV},  \quad \lambda_\chi = 0.38,  \quad \lambda_\psi = 0.40,  \quad \lambda_\eta = 0.51, \notag \\
& y_2 = y_4=1.6 \times 10^{-3},   \quad  y_1 = y_3 = 9.9 \times 10^{-3},
\end{align}
while for Benchmark 2:
\begin{align}
& v_s = 1666~\text{GeV},  \quad \lambda_\chi = 0.25,  \quad \lambda_\psi = 0.27,  \quad \lambda_\eta = 0.34, \notag \\
& y_2 = y_4=1.6 \times 10^{-3},   \quad  y_1 = y_3 = 9.9 \times 10^{-3}.
\end{align}
Strictly speaking, with the above parameters the masses $m_{N_2}$, $m_{E_1}$ are not exactly degenerate and equal to $m_\psi$, e.g. for Benchmark 1, using expressions (\ref{M1})--(\ref{M3}), we get mass shifts $\Delta_N = 5.6$ MeV, $\Delta_E = 34$ MeV, which are negligible for LHC phenomenology.

In these scenarios we have the following decays of the heavy leptons:
\begin{itemize}
\item $N_2 \to N_1 f \bar f$, where $f$ is any fermion except the top quark. These decays are mediated by an off-shell $Z$ boson, and the final state with $f=b$ receives a small contribution from Higgs boson exchange. The decays of interest, $N_2 \to N_1 \ell^+ \ell^-$ have a branching ratio of 3.9\% for $\ell=e,\mu$ and 3.6\% for $\ell=\tau$.

\item $E_1 \to N_1 f \bar f'$, with $f \bar f'=d \bar u, s \bar c, \ell^- \nu$. The $Z' \to E_1 E_1$ decay produces signals with zero, one or two soft leptons plus soft jets and small missing energy. Clearly, there is little hope for such signals.

\item The heavier charged lepton can in principle decay $E_2 \to E_1 Z$, $E_2 \to N_2 W$, $E_2 \to N_1 W$ or $E_2 \to E_1 H$. The partial widths are proportional to $(\sin 2 \theta^E)^2$, $(\cos \theta^N \times \cos \theta^E)^2$, $(\sin \theta^N \times \cos \theta^E)^2$ and $(\sin 2\theta^E \times \cos 2 \theta^E)^2$, respectively. Hence, for 
$\theta^N\simeq 0$ and $\theta^E \simeq \pi/2$,  the $E_2 \to N_1 W$ decay is suppressed with respect to the others. The other three are sizeable if they are kinematically allowed, although $E_2 \to N_2 W$ is typically the dominant one. For our benchmarks it turns out that this is in fact the only kinematically allowed mode, so it has a nearly 100\% branching ratio.
\end{itemize}
The decays to leptons of different flavour, i.e. $Z' \to N_1 N_2$, $Z' \to E_1 E_2$, are very suppressed in the scenarios with small mixings considered here. Note also that there is a contribution to four-lepton signals from $Z' \to E_2 E_2$, which we take into account in our simulations.

\section{Discovery potential}
\label{sec:6}

 We consider five sets of centre-of-mass (CM) energies and luminosities: (i) Run 2 of the LHC, with 13 TeV and a luminosity $L=150$ fb$^{-1}$; (ii) Run 3 of the LHC, with 14 TeV and $L=300$ fb$^{-1}$; (iii) the high-luminosity LHC (HL-LHC), with $L= 3$ ab$^{-1}$ at 14 TeV; (iv) a high-energy upgrade of the LHC (HE-LHC), with $L= 15$ ab$^{-1}$ at 27 TeV; (v) a future circular collider (FCC) with 
$L= 30$ ab$^{-1}$ at 100 TeV.

We generate our signals (in the above benchmark scenarios) and the backgrounds using {\scshape MadGraph5}~\cite{Alwall:2014hca}. For the signal processes the relevant Lagrangian is implemented in {\scshape Feynrules}~\cite{Alloul:2013bka} and interfaced to {\scshape MadGraph5} using the universal Feynrules output~\cite{Degrande:2011ua}. Tau leptons are included in all processes. Hadronisation and parton showering are performed with {\scshape Pythia~8}~\cite{Sjostrand:2007gs} and detector simulation using {\scshape Delphes 3.4}~\cite{deFavereau:2013fsa} using the configuration for the ATLAS detector for LHC Runs 2 and 3. For HL-LHC and HE-LHC we use a card corresponding to the expected performance of the upgraded ATLAS/CMS detectors~\cite{card}, and for the FCC the corresponding card. In all cases, we relax the isolation requirement on charged leptons because in the signal the leptons from each $N_2$ decay are relatively close. This is an approximation, at the level of fast simulation, of the loose lepton isolation criteria that are applied in experimental searches with full control over the details of the lepton definition and reconstruction. 

As pre-selection criteria, we require that events have exactly four leptons (electrons or muons), in two opposite-sign same-flavour pairs. We set a lower cut $p_T^\ell \geq$ 10 GeV for all leptons and for LHC Runs 2 and 3 we select events that fulfill at least one of the following criteria from the Run 2 ATLAS trigger menu~\cite{trigger}:
\begin{itemize}
\item one electron with $p_T \geq 27$ GeV;
\item one muon with $p_T \geq 27$ GeV;
\item two electrons with $p_T \geq 18$ GeV;
\item two muons with $p_T \geq 15$ GeV;
\item one muon with $p_T \geq 23$ GeV and another muon with $p_T \geq 9$ GeV;
\item one muon with $p_T \geq 25$ GeV and another muon with $p_T \geq 8$ GeV;
\item one electron with $p_T \geq 17$ GeV and two electrons with $p_T \geq 9$ GeV;
\item three muons with $p_T \geq 6$ GeV.
\end{itemize}
The effect of the trigger on the signal efficiencies is minimal. For LHC upgrades the multilepton triggers are planned to even lower their thresholds; moreover, a low $p_T$ four-lepton trigger consumes very little bandwith (because the SM four-lepton background is quite small) and could easily be implemented. We therefore do not apply any trigger requirement for LHC upgrades and for the FCC, besides the common requirement of $p_T^\ell \geq$ 10 GeV for all leptons. For future colliders, the computing capabilities will have to match the high output rates in other processes with much larger cross sections than four-lepton production. In that case, a four-lepton trigger with low threshold (such as $p_T \geq 10$ GeV) will be of little extra bandwidth and will allow to record the signals discussed.

The main irreducible backgrounds to our signals are four lepton production $pp \to 4\ell$, mediated by off-shell $Z$ bosons and photons, Higgs production with decay $H \to ZZ^*$ and five lepton production $pp \to 5 \ell + \nu$, also involving off-shell $Z$ bosons and photons. Note that a much larger source of four leptons is for example $t \bar t$ production in the dilepton decay mode, with the two additional leptons from $b$ quark decays. This and other backgrounds can be quite reduced by requiring that the additional energy within a small cone, typically of radius $R=0.2$ around the lepton, amounts to a small fraction of the lepton energy~\cite{Aaboud:2018fvk}. (Contributions to the isolation cones from other leptons are subtracted before applying the requirements.) With these isolation criteria, $t \bar t$ and $b \bar b$ are an order of magnitude below the former irreducible backgrounds.  Since the tools to deal with this type of backgrounds are not available at the level of fast simulation and they are quite smaller than the irreducible ones, we do not include them in our calculations.

\begin{figure}[t]
\begin{center}
\begin{tabular}{cc}
\includegraphics[height=5.2cm,clip=]{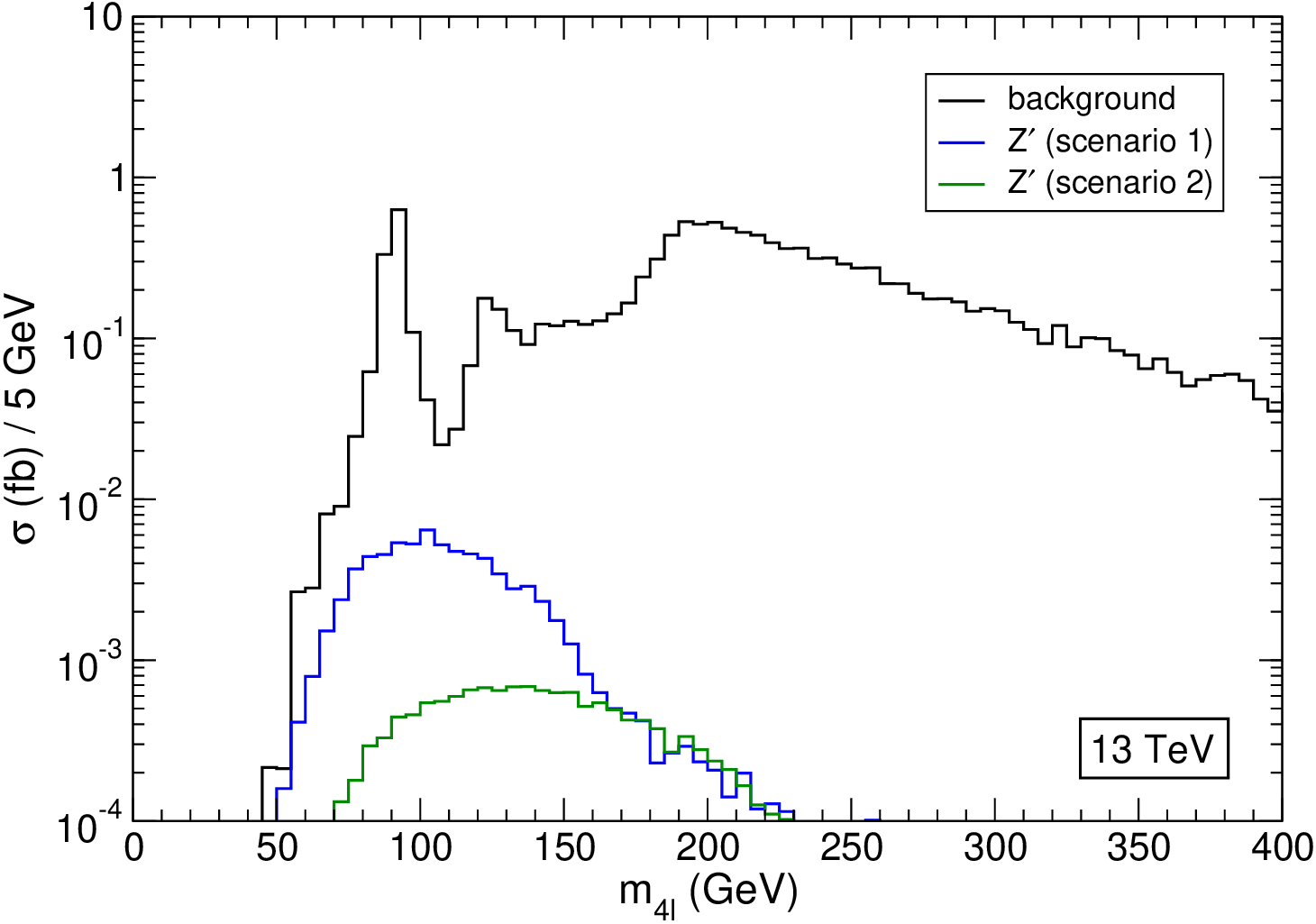} 
&  \includegraphics[height=5.2cm,clip=]{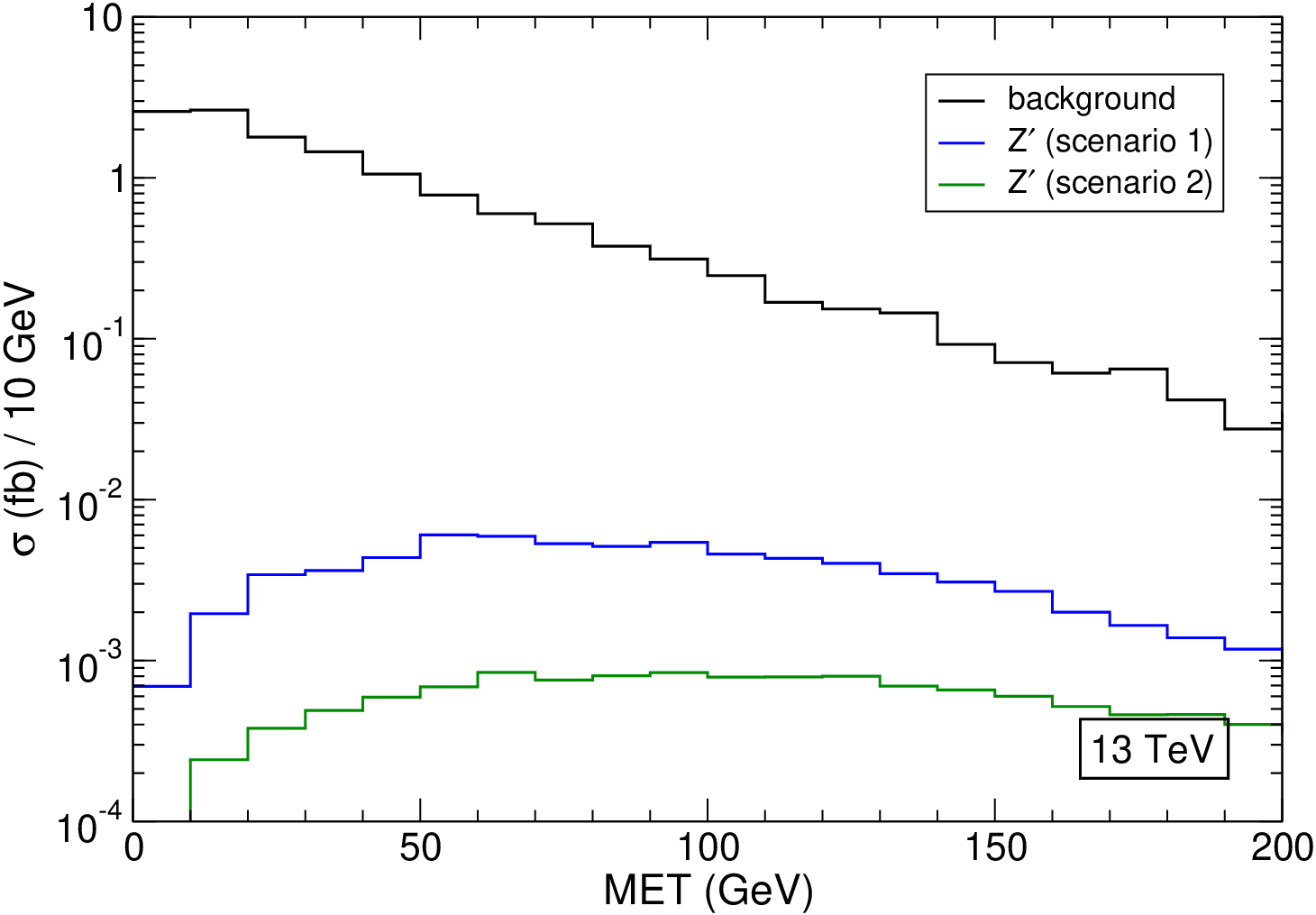} 
\end{tabular}
\caption{Four-lepton invariant mass (left) and missing energy (right) for the signal and the SM background, for a CM energy of 13 TeV and the two benchmark scenarios defined in Eqs(\ref{ec:Benchmark1}, \ref{ec:Benchmark2}).}
\label{fig:m4l}
\end{center}
\end{figure}

We point out that this signal shares some features with the exotic Higgs decays $H \to XX \,,\, X \to \ell^+ \ell^-$, with $X$ a new light boson or a pseudo-scalar, which is searched for at the LHC~\cite{Aad:2015sva,Aaboud:2018fvk}. We show in figure~\ref{fig:m4l} (left) the four-lepton invariant mass for the signals and the SM background, for a CM energy of 13 TeV and the two benchmark scenarios defined in Eqs.~(\ref{ec:Benchmark1}), (\ref{ec:Benchmark2}). The four-lepton invariant mass is in the range not far from the Higgs mass and of course it does not display a peak. Notice the background peak at $M_Z$, when two of the leptons are emitted in the radiative decay of an on-shell $Z$ boson, and the smaller peak at $M_H$, caused by $H \to ZZ^*$.
On the right panel we show the missing energy distributions. The latter has some discrimination power between the signals and the backgound but for simplicity we do not use it as the improvement on the signal significance is small.

An excellent discrimination between the signals and the background is achieved by using the minimum sum of dilepton invariant masses $\Sigma m_{\ell \ell}$, defined as follows. Among the possible pairings of opposite-sign same-flavour pairs $(\ell_1^+ \ell_1^-)$, $(\ell_2^+ \ell_2^-)$  --- there is only one pairing in $e^+ e^- \mu^+ \mu^-$ events, but there are two if all the leptons have the same flavour --- we select the one that minimises the sum of the two invariant masses $m_{\ell_1^+ \ell_1^-} + m_{\ell_2^+ \ell_2^-}$. This minimum is $\Sigma m_{\ell \ell}$. 
\begin{figure}[t]
\begin{center}
\begin{tabular}{cc}
\includegraphics[height=5.5cm,clip=]{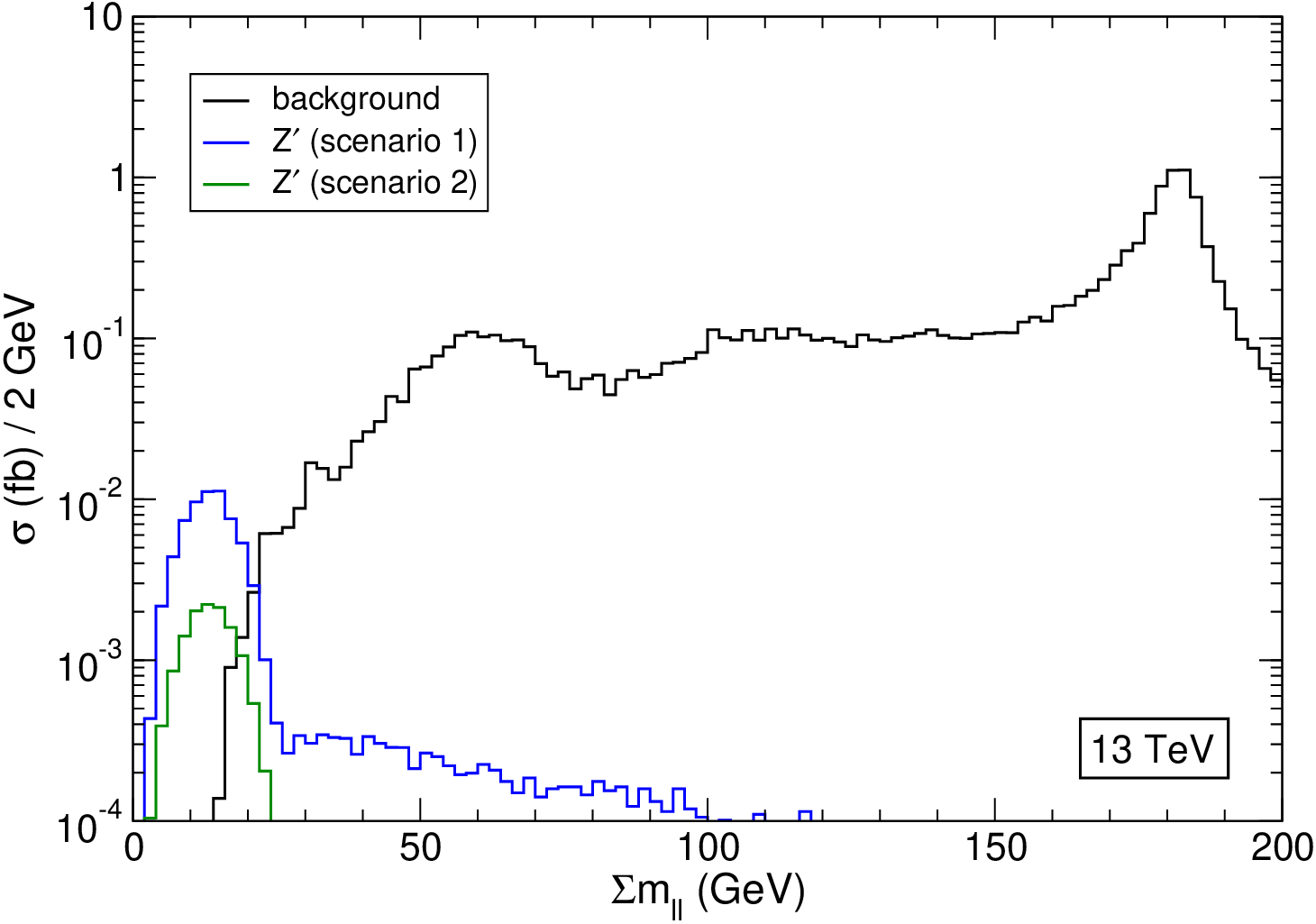} &
\includegraphics[height=5.5cm,clip=]{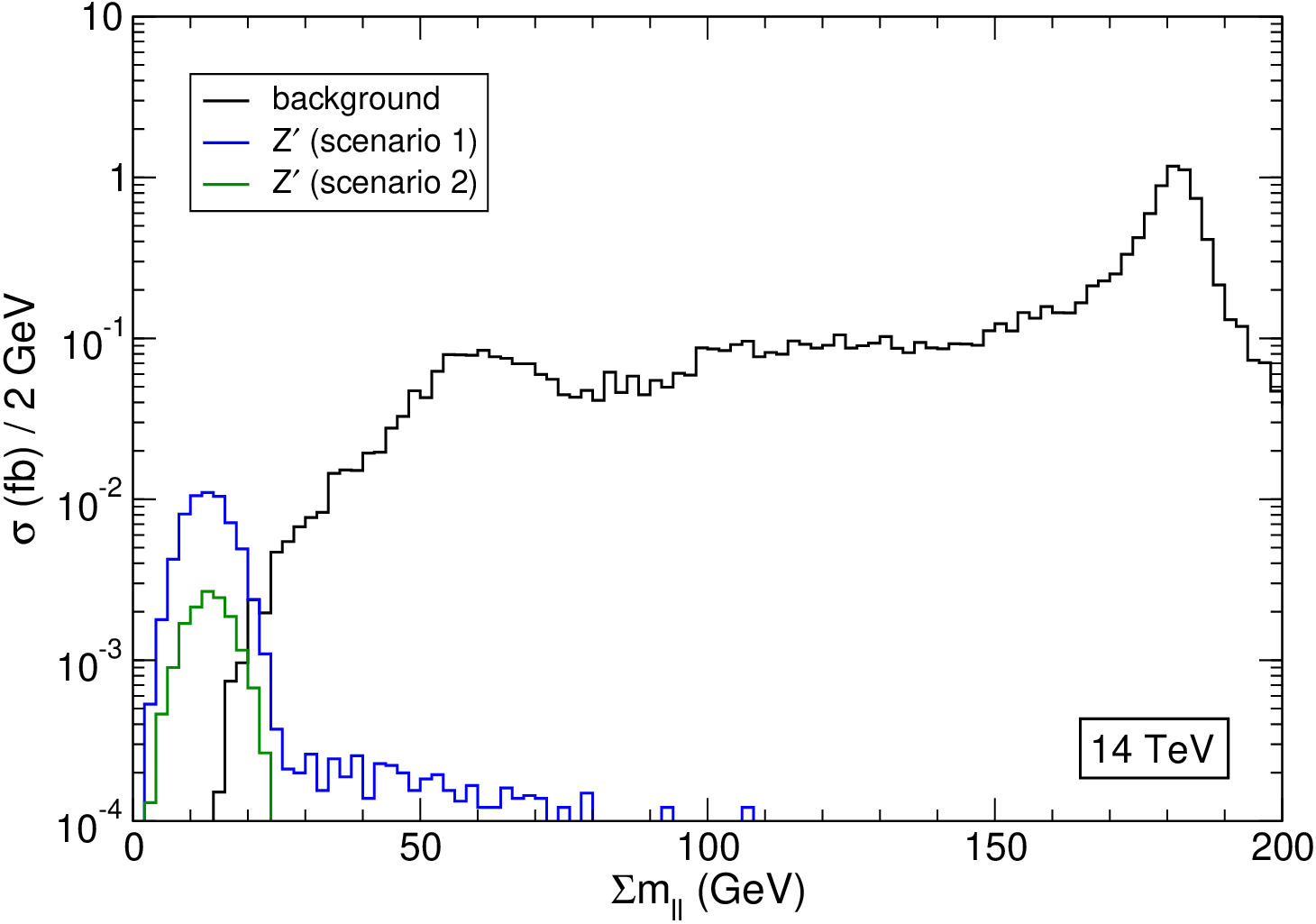}  \\
\includegraphics[height=5.5cm,clip=]{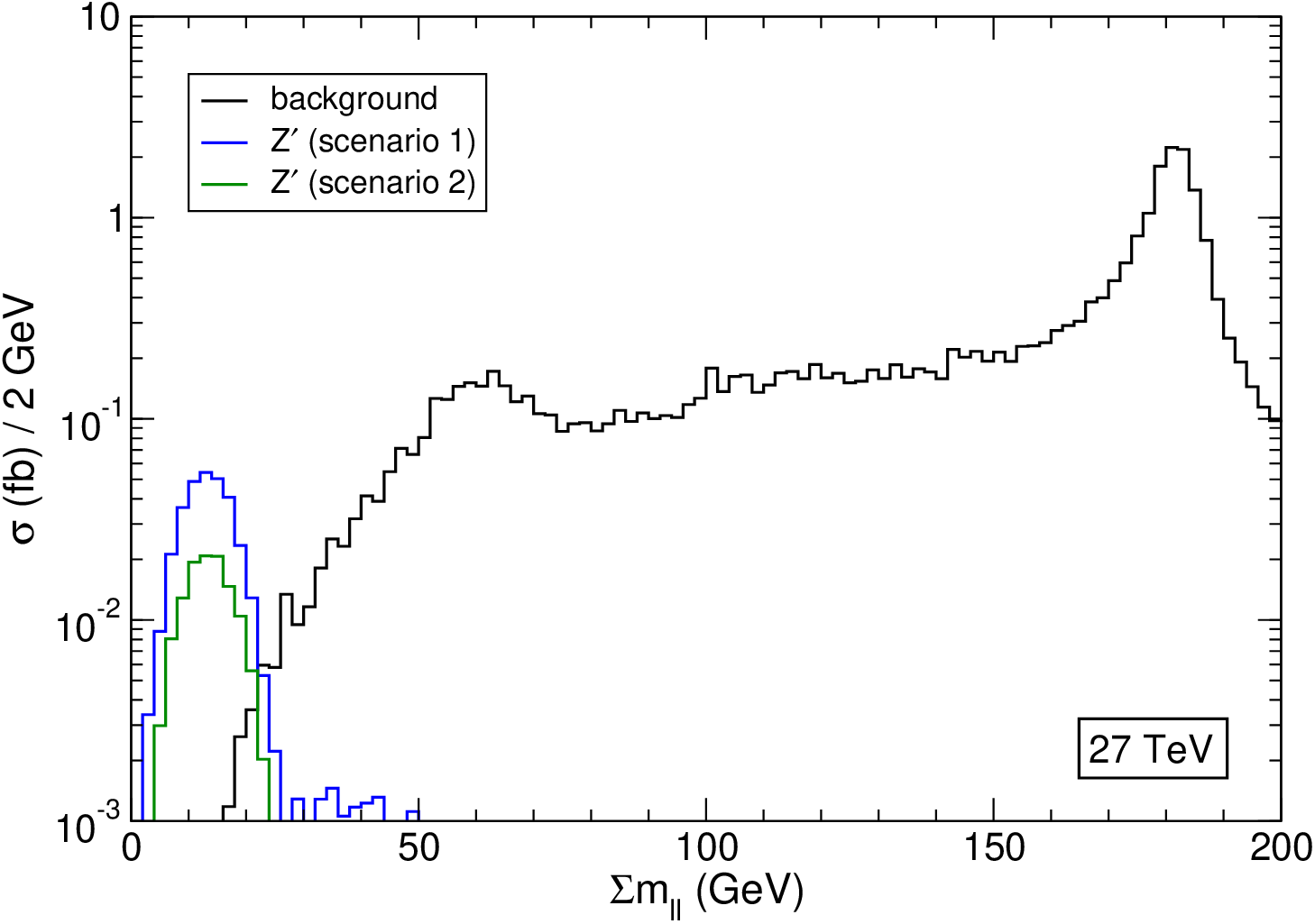}  &
\includegraphics[height=5.5cm,clip=]{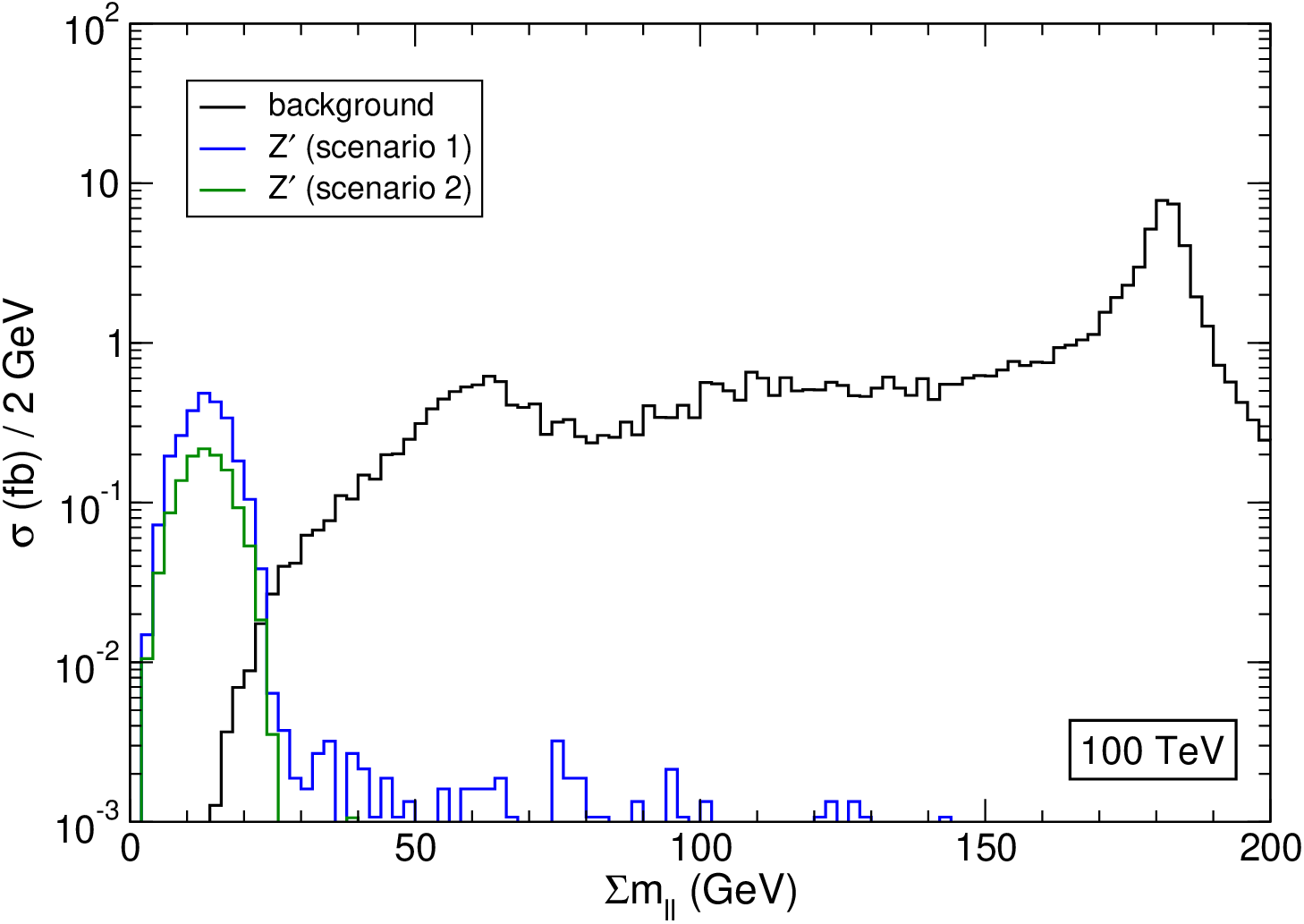} 
\end{tabular}
\caption{Kinematical distribution of the minimum sum of dilepton invariant masses $\Smll$ for signals and backgrounds, for four CM energies.}
\label{fig:Smll}
\end{center}
\end{figure}
For the dominant process giving four leptons, $Z' \to N_2 N_2$, it turns out that $\Sigma m_{\ell \ell}\leq 2 (m_{N_2} - m_{N_1})$, since there is at least one pairing, the one corresponding to leptons with the same mother particle, which fulfills such inequality, see eq.~(\ref{mll}). Therefore we expect an accumulation of the four-lepton signal in that range of small $\Sigma m_{\ell \ell}$. This is fortunate since that is precisely the region where the background is less important. Actually, the only relevant background is four-lepton production, and the rest are two orders of magnitude below. (As mentioned, other backgrounds with leptons from top / bottom quark decays are expected to be unimportant.) All this is illustrated in figure~\ref{fig:Smll}, which shows the distributions of signals and backgrounds for Run 2 (top, left), HL-LHC (top, right), HE-LHC (bottom, left) and FCC (bottom, right). 

\begin{table}[t]
\begin{center}
\begin{tabular}{lcccc}
& \multicolumn{2}{c}{Run 2} &  \multicolumn{2}{c}{Run 3} \\
& \multicolumn{2}{c}{\small pre-sel. / selection} & \multicolumn{2}{c}{\small pre-sel. / selection} \\
$Z' \to N_2 N_2$ S1 & 0.050 & 0.049 & 0.061 & 0.060 \\
$Z' \to E_2 E_2$ S1 & 0.025 & 0.014 & 0.032 & 0.018 \\
$Z' \to N_2 N_2$ S2 & $9.1 \times 10^{-3}$ & $8.9 \times 10^{-3}$& 0.013 & 0.012 \\
$Z' \to E_2 E_2$ S2 & $5.3 \times 10^{-3}$ & $3.4 \times 10^{-3}$ & $7.2 \times 10^{-3}$ & $4.7 \times 10^{-3}$ \\
$4\ell$ & 13.1 & $5.2 \times 10^{-3}$ & 14.1 & $5.3 \times 10^{-3}$ \\
$H \to ZZ^*$ & 0.218 & $1.7 \times 10^{-5}$ & 0.243 & $3.0 \times 10^{-5}$ \\
$5\ell$ & 0.046 & $4.8 \times 10^{-6}$ & 0.050 & $4.7 \times 10^{-6}$ \\
\end{tabular}
\vspace{0.5cm}

\begin{tabular}{lcccccc}
& \multicolumn{2}{c}{HL-LHC} &  \multicolumn{2}{c}{HE-LHC}  &  \multicolumn{2}{c}{FCC} \\
& \multicolumn{2}{c}{\small pre-sel. / selection} &  \multicolumn{2}{c}{\small pre-sel. / selection} & \multicolumn{2}{c}{\small pre-sel. / selection} \\
$Z' \to N_2 N_2$ S1 & 0.049 & 0.048 & 0.241 & 0.236 & 2.06 & 2.02 \\
$Z' \to E_2 E_2$ S1 & 0.021 & 0.013 & 0.107 & 0.064 & 0.525 & 0.436 \\
$Z' \to N_2 N_2$ S2 & 0.011 & 0.010 & 0.088 & 0.086 & 0.957 & 0.939 \\
$Z' \to E_2 E_2$ S2 & $5.3 \times 10^{-3}$ & $3.7 \times 10^{-3}$ & 0.044 & 0.031 & 0.296 & 0.249 \\
$4\ell$ & 12.5 & $4.2 \times 10^{-3}$ & 23.2 & $7.7 \times 10^{-3}$ & 71.4 & $0.021$\\
$H \to ZZ^*$ & 0.204 & $2.3 \times 10^{-5}$ & 0.603 & $3.0 \times 10^{-5}$ & 3.39 & $4.2 \times 10^{-4}$ \\
$5\ell$ & 0.042 & $1.7 \times 10^{-6}$ & 0.092 & $1.5 \times 10^{-5}$ & 0.257 & $1.8 \times 10^{-5}$ \\
\end{tabular}
\end{center}
\caption{Cross sections (in fb) of the different signals and backgrounds at LHC and its upgrades at the pre-selection and final selection. The signal labels S1, S2 refer to benchmark scenarios 1 and 2, defined in Eqs(\ref{ec:Benchmark1}, \ref{ec:Benchmark2}). The background labels $4\ell$ and $5\ell$ correspond to four- and five-lepton production as described in the text.}
\label{tab:xs}
\end{table}

Two distinct regions for the signals can be distinguished. A region of small $\Smll$ arises from $Z' \to N_2 N_2$, with a small contribution from $Z' \to E_2 E_2 \to N_2 W N_2 W$, with hadronic $W$ decay. As expected, this accumulation of signal occurs at $\Sigma m_{\ell \ell}\leq 2( m_{N_2} - m_{N_1}) $, which equals 28 GeV in these examples. Besides this region, there are 
signal tails caused by $Z' \to E_2 E_2 \to N_2 W N_2 W$ when one or both $W$ bosons decay into electrons or muons. In this decay chain the $e/\mu$ resulting from $W$ decay typically have larger $p_T$ than the ones from $N_2 \to N_1 \ell \ell$ and their pairing with other leptons does not result in small $\Smll$. Concerning the background, the SM production of four leptons peaks at $2 M_Z$, as expected, and it is three orders of magnitude smaller at the signal region of small $\Smll$. This makes the $\Sigma m_{\ell \ell}$ variable a very convenient one to bring to light a `compressed' spectrum, as the one expected in co-annihilation regimes, provided the co-annihilating particle may decay with two leptons in the final state.

We require as selection criterium $\Smll \leq 22$ GeV in all cases. The breakdown of signal and background cross sections for the different processes considered is given in table~\ref{tab:xs}.\footnote{The cross sections at 14 TeV for Run 3 and HL-LHC are not equal because of the different detection efficiencies and energy resolutions in the Delphes cards for the ATLAS and expected HL-LHC detector; in particular, the latter card is based on projections for the CMS detector.} The extra contribution to the four-lepton signal from $Z' \to E_2 E_2$ amounts to 40\%--50\%; however, at the region of small $\Smll$ it is smaller, around 30\%.

With the number of signal ($S$) and background ($B$) events obtained we compute the expected signal significances for Runs 2 and 3, using Poisson statistics. These numbers are collected in table~\ref{tab:SB1}. We do not include any systematic uncertainty, as the statistical one is clearly dominant for a background of less than two events. (For future colliders the background is larger than a handful of events but still one can use the sideband for a precise normalisation of the background; the Monte Carlo predictions for four-lepton production are reliable since it is an electroweak process.)
For LHC upgrades it is not sensible to report the relative sensitivities in terms of signal significances $n_\sigma$ for a fixed coupling  --- in a scenario that would have been discovered with $5\sigma$ well before anyway. Instead, we give in table~\ref{tab:SB2} the couplings for which the signals could be seen with $5\sigma$ significance for the two benchmarks. (Poisson statistics are still used for HL-LHC with a background of $12$ events; for HE-LHC and FCC we use the Gaussian approximation.) Given the fact that this process has tiny background, the potential of future colliders is really impressive: couplings at the few percent level could be probed.

\begin{table}[t]
\begin{center}
\begin{tabular}{lcccccc}
& \multicolumn{2}{c}{Run 2} & \multicolumn{2}{c}{Run 3} \\
$B$ &  \multicolumn{2}{c}{0.78} &  \multicolumn{2}{c}{1.60} \\
$S$ (S1) & 9.3 & $5.7\sigma$ & 23.1 & $8.3\sigma$ \\
$S$ (S2) & 1.9 & $1.3 \sigma$ & 5.1 & $2.7\sigma$ 
\end{tabular}
\end{center}
\caption{Expected number of signal ($S$) and background ($B$) events, and statistical significance ($n_\sigma$) of the signal, for LHC Runs 2 and 3. The $Z'$ coupling is set as $g_{Z'} Y'_q = 0.2$.}
\label{tab:SB1}
\end{table}

\begin{table}[htb]
\begin{center}
\begin{tabular}{lcccc}
& HL-LHC & HE-LHC & FCC  \\ 
S1 & 0.069 & 0.022 & $8.3 \times 10^{-3}$ \\
S2 & 0.14 & 0.035 & $0.012$ \\
\end{tabular}
\end{center}
\caption{Coupling $g_{Z'} Y'_q$ for which a signal can be seen with $5\sigma$ significance at future colliders.}
\label{tab:SB2}
\end{table}

Finally, let us comment about the production of dark lepton pairs mediated by off-shell $Z/\gamma$ or $W$ bosons. As it is clear from the analogy with supersymmetric compressed spectra, these signals are almost invisible as they produce small missing energy and very soft leptons or jets. For definiteness, we can quantify this statement for the masses and mixing angles used in the previous benchmarks, c.f. (\ref{ec:Benchmark1}), (\ref{ec:Benchmark2}) in three representative cases.
\begin{itemize}
\item $pp \to Z \to N_2 N_2 \to N_1 \ell^+ \ell^- \, N_1 \ell^+ \ell^-$. The cross section, summing $\ell = e,\mu$, is 0.18 fb at 13 TeV. However, as seen in figure~\ref{fig:parton1} the leptons are quite soft, and requiring that the four of them have transverse momentum $p_T^\ell \geq 10$ GeV (at the parton level) yields a suppression by a factor of $0.012$, resulting in a cross section of $2.1 \times 10^{-3}$ fb. For comparison, $Z'$-mediated $N_2 N_2$ production with $p_T^\ell \geq 10$ GeV has cross sections of 0.078 fb for $M_{Z'} = 2$ TeV and $0.014$ fb for $M_{Z'} = 3$ TeV.
\item  $pp \to Z/\gamma \to E_1^+ E_1^- \to N_1 \ell^+ \nu \, N_1 \ell^- \nu$. The cross section with $\ell = e,\mu$, is 1.2 fb at 13 TeV. Requiring $p_T^\ell \geq 10$ GeV reduces the signal by a factor of $0.073$, yielding a cross section of 0.85 fb. For comparison, $W^+ W^- \to \ell^+ \nu \ell^- \nu$ with $p_T^\ell \geq 10$ GeV has a cross section of 2.61 pb, more than three orders of magnitude larger. Because the $E_1^+ E_1^-$ signal is kinematically similar to Drell-Yan $N_2 N_2$ production in figures~\ref{fig:parton1} and \ref{fig:parton2}, without very distinctive features, it is likely unobservable.
\item $pp \to W^\pm \to E_1^\pm N_2 \to N_1 \ell^\pm \nu \, N_1 \ell^+ \ell^-$. The cross section with $\ell=e,\mu$ is 1.9 fb at 13 TeV. Requiring that the three charged leptons have $p_T^\ell \geq 10$ GeV amounts to a suppresion factor of $0.027$, reducing the cross section to $0.052$ fb. The cross section of the $\ell^\pm \nu \ell^+ \ell^-$ background can be reduced to 6.7 fb by requiring that the invariant mass of the opposite-sign same-flavour pair produced from the $Z/\gamma$ is smaller than 14 GeV. Still, this is two orders of magnitude above the signal, which is unobservable.
\end{itemize}
Therefore, one can see that, as a generic feature, the direct production of dark lepton pairs gives signals that are quite difficult to see because of the kinematics of the compressed spectrum. When these leptons have masses above few hundreds  of GeV the cross sections are also quite small, and the signals are unobservable. Final states where the dark leptons decay into soft jets are even more invisible.

\section{Summary and discussion}
\label{sec:7}

WIMP models of thermal dark matter require an appropriate annihilation of the latter in the early universe, and thus interactions with the SM particles. The most obvious of such interactions are those mediated by a Higgs or a $Z$ boson (Higgs and $Z$ portals). However, this possibility is under strong experimental pressure, essentially from direct detection constraints. Then, one of the best motivated and popular scenarios of dark matter is when those interactions occur through a $Z'$ boson ($Z'$ portal). In order to avoid strong constraints from direct detection experiments and  dilepton production at the LHC, it is highly convenient that the $Z'$ couplings are both leptophobic and axial (to either quarks or dark matter).
This framework has been much explored in the literature, but usually in the context of `simplified dark matter models', where only the dark matter particle plus the mediator, $Z'$, are considered. This leads to very characteristic signals already searched for at the LHC, such as mono-Higgs~\cite{Aaboud:2017yqz,Sirunyan:2019zav}, mono-top~\cite{Aaboud:2018zpr,Sirunyan:2019gfm}, mono-$Z/W$~\cite{Aaboud:2018xdl,Sirunyan:2017qfc,Sirunyan:2017jix} and mono-jet~\cite{Aaboud:2017phn,Sirunyan:2017jix} production. The common feature of these signals is the production of a SM particle together with large missing energy resulting from the undetected dark matter particle. Unfortunately, no positive signal has shown up in any of these experimental searches, up to date.

However, these simplified models are in fact {\em not} minimal, since they present various theoretical inconsistencies, in particular the lack of anomaly cancellation. The latter requires to extend the dark sector with at least three extra fermions, a $SU(2)$ doublet and a $SU(2)$ singlet, both with non-vanishing hypercharge and extra hypercharge (the one associated to the extra $U(1)$ gauge group) \cite{Duerr:2014wra,Ellis:2018xal,Caron:2018yzp}.
The presence of these extra states affects both the phenomenology of dark matter both at the early universe, due to possibility of co-annihilations, and at the LHC, as novel dark matter signals may appear. The goal of this paper has been to explore this new phenomenology, with the focus on its possible detection at the LHC and future colliders.

We have studied a particularly clean signal consisting of four charged leptons, with (perhaps surprisingly) small missing energy, which arises from the cascade decay  $Z' \to N_2 N_2 \to N_1 \ell^+ \ell^- \, N_1 \ell^+ \ell^-$. Its most salient feature is the presence of two opposite-sign same-flavour lepton pairs of low invariant mass. Because the main source of four leptons in the SM --- barring other sources that produce them close to jets such as $b$ quark decays --- is on-shell $ZZ$ production, the backgrounds for such signal are tiny. 

Searches for this type of dark matter scenarios can be performed in four-lepton events by using the discriminant variable $\Smll$, that is, the minimum sum of invariant masses of opposite-sign same-flavour pairs. The signals can be spotted as an excess at the low-$\Smll$ region. We have verified that such events can be triggered already at the LHC Run 2, and the expected backgrounds have quite different kinematical features, so that even if the signal has a small cross section, as it corresponds to the production of a TeV scale $Z'$, it could be seen for reasonable values of the model parameters.

Current searches for exotic Higgs decays $H \to XX \to 4\mu$ could also be extended to have sensitivity to the signals introduced here. For example, ref.~\cite{Aaboud:2018fvk} has a dedicated analysis for $1~\text{GeV} \leq M_X \leq 15~\text{GeV}$, considering the four-muon final state. However, the analysis focuses on a narrow four-lepton invariant mass window $m_{4\ell} \in [120,130]$ GeV around the Higgs boson mass --- adequate for a search of Higgs exotic decays --- that unfortunately removes most of our signal, as it can readily be observed in figure~\ref{fig:Smll}. In addition to this cut, the event reconstruction is done by assuming the kinematics of the decay $H \to XX$, and is sub-optimal for the signals addressed here.

Besides the production of an excess in the low-$\Smll$ region, which is the common feature, other characteristics of the signal depend on the model parameters. The full exploration of the  relevant five-dimensional parameter space is cumbersome,  but one can easily figure out, from the results in section~\ref{sec:4}, the behaviour for parameters other than those considered in the detailed simulation in  section~\ref{sec:6}.
\begin{itemize}
\item[(i)] $Z'$ mass: for heavier $Z'$ the cross section is obviously smaller; the leptons are produced with higher transverse momentum and therefore the efficiency for event selection is larger; the missing energy is also larger the heavier the $Z'$ is. 
\item[(ii)] $N_1, N_2$ masses: the lepton $p_T$ are proportional to the relative mass splitting $(m_{N_2}-m_{N_1})/m_{N_2}$, so a smaller mass difference or heavier $N_2$ makes the signals harder to see. It should be noticed here that the heavier $N_1$, the smaller the relative splitting must be in order to enhance the co-annihilation effects. In addition, heavier $N_1$, $N_2$ also implies smaller $Z' \to N_2 N_2$ branching ratio.
\item[(iii)] $E_2$ mass: the signal receives a small contribution if $E_2$ is close to $N_1$, $N_2$, but otherwise the influence is moderate.
\item[(iv)] Coupling: the signal cross section scales with $g_{Z'}^2$ but this coupling cannot be arbitrarily large, since one has the limit $g_{Z'} Y'_q \lesssim 0.3$ from dijet resonance searches for the $Z'$ masses considered.
\end{itemize}
As a final comment, let us remark that in this paper we have considered benchmark scenarios where the extra scalar(s) $S$ necessary to provide the $Z'$ and dark lepton masses do not play any role in $Z'$ decays, by taking them heavy. Conversely, in ref.~\cite{Aguilar-Saavedra:2019adu} the dark leptons where assumed heavy, to concentrate on the phenomenology of the $Z'$ boson cascade decays into extra scalars. The perhaps more natural (and quite more complex) situation is to have scalars and dark leptons with mass of the same order, so that the new scalars can decay into dark leptons and vice versa. The analysis of this type of scenarios and their possible collider signals deserves further investigation. 

\section*{Acknowledgements}  This work has been supported by Spanish Agencia Estatal de Investigaci\'on through the grant `IFT Centro de Excelencia Severo Ochoa SEV-2016-0597' and by MINECO projects FPA 2016-78022-P, FPA 2016-78220-C3-1-P and FPA 2017-85985-P (including ERDF), by the Spanish Red Consolider MultiDark FPA 2017-90566-REDC, and by European Union 
through the Elusives ITN (Marie Sklodowska-Curie grant agreement No 674896).

\end{document}